\documentstyle[12pt,bezier]{article}
\def\rpp#1#2#3#4{\phi_{#1}^{#2}\,{}^{(#3)}(#4)}
\def\f#1#2{{\cal F}^{(#1)}(\nu,#2)}
\def\q#1#2{Q^{(#1)}(#2)}
\def\n#1{J^0_{#1}}
\def\e#1{J^+_{#1}}
\def\j-#1{J^-_{#1}}
\def\hv#1#2{|#1\rangle ^{(#2)}}
\def\F#1#2{{\cal F}_{#1}^{(#2)}}
\def\l#1{\omega^{\dagger}_{#1}}
\def\dom#1{\omega^{\dagger}(#1)}
\def\o#1{\omega_{#1}}
\def\om#1{\omega(#1)}
\def\a#1{a_{#1}}
\def\vp#1{\varphi(#1)}
\def\rpf#1#2#3{\phi_{#1}^{(#2)}(#3)}
\def\pf#1#2{\phi^{#1}(#2)}
\def\cvc{\langle vac|}
\def\cvac{\langle 0|}
\def\vc{|vac\rangle}
\def\vac{|0\rangle}
\def\NP#1#2{{\it Nucl.Phys.} {\bf B#1} (#2)}

\def\CMP#1#2{{\it Commun.Math.Phys.} {\bf #1} (#2)}
\vspace{2cm}
\title{{ \bf On bosonization of 2d conformal field theories }}
\author
{Oleg Andreev\thanks{e-mail: andreev@ifh.de}\,\,\thanks{supported by DFG}
\\ Humboldt-Universit\"{a}t zu Berlin, \\
Mathematisch-Naturwissenschaftliche Fakult\"{a}t I,\\Institut f\"{u}r Physik,
Invalidenstra\ss e 110,  \\ \vspace{1.5cm} 10099 Berlin, Germany \\ Boris
Feigin
\\ Landau Institute for Theoretical Physics, Kosygina 2,  \\ Moscow 117334,
Russia}

\date{}
\begin{document}

\maketitle

\begin{abstract}
\vspace{0.7cm}
We show how bosonic (free field) representations for so-called degenerate
conformal theories are built by singular vectors in Verma modules. Based on
this
construction, general expressions of conformal blocks are proposed. As an
example we describe new modules for the $SL(2)$ Wess-Zumino -Witten model.
They are, in fact, the simplest non-trivial modules in a full set of bosonized
highest weight representations of $\hat {sl}_2$ algebra. The Verma and Wakimoto
modules appear as boundary modules of this set. Our construction also yields a
new kind of bosonization in 2d conformal field theories.

\end{abstract}
\vspace{-18.5cm}
\hspace{11.5cm}
\begin{tabular}{ll}
HUB-IEP-94/3 \\

hep-th/9403081
\end{tabular}

\newpage

\section{Introduction}

Conformal field theory plays a crucial role in string theory and 2d statistical
mechanics [1]. In order to solve the theory it is important to find correlation
functions for a set of basic conformal operators (primary fields).

The free field representation provides in principle a powerful method to obtain
the correlators and to compute the operator algebra (OA) of the primary fields
[2-10].  Examples of models which have been solved by this approach are given
by
the minimal conformal models and the $SL(2)$ Wess-Zumino-Witten (WZW) model
[3,9,10]. The basic idea of the free field representation is to construct the
Hilbert space of theory by means of Fock modules of free fields. However the
danger of unphysical states remains present: a careful analysis of the
BRST-like
complex of Fock modules is needed.

At present a technique of building the free field representation (generators of
symmetry, primary fields, correlators, BRST-like complexes etc.) of some
conformal field theories is developed [3,4,6-8,10,11]. We shall argue that the
above technique has not yet been fully realized. The missing points are the
full
set of representations for the primary operators as well as a proper definition
for conformal blocks, BRST complexes etc.

Throughout this paper , we shall always consider only degenerate conformal
theories. Note that they have important applications in physics, for example in
a describtion of critical fluctuations in a variety of
statistical models [1,12].

The outline of the paper is as follows.

The main body of this work is presented in section 2. Since Kac-Moody algebras
play a central role in 2d conformal field theories (most known examples of
conformal models can be understood in terms of WZW model) we discuss the
construction of the full set of the representations for the primary fields in
the WZW model for simple Lie algebra $g$. As a result we obtain a general
representation of conformal blocks of this model on the plane.

To demonstrate the approach we shall concentrate below on the $SL(2)$ WZW
model.
The generalization to other models is straightforward.

In section 3, the brief review of the Wakimoto representation for the $SL(2)$
WZW model is given. In particular, the resolution of an
irreducible module (Hilbert space) of the algebra $\hat {sl}_2$ in terms of
Fock
spaces is sketched.

Section 4 provides the construction of the first nontrivial representation,
the so-called Dotsenko representation, for the $SL(2)$ WZW model. The structure
of the Dotsenko modules over $\hat {sl}_2$ and the resolution of an irreducible
module
is also briefly discussed. As an example the two-point function of the primary
fields is computed.

In section 5 we present the second nontrivial representation. We analyze its
structure over $\hat {sl}_2$ and give the resolution of an irreducible module
in
terms of Fock spaces. Also the two-point function of the primary
fields is computed.

The last, section 6, contains some conclusions and speculations.

In the appendices we give technical details which are relevant for the explicit
construction of the nontrivial bosonized representations of the $SL(2)$ WZW
model.

\section{The method}

Let $|j\rangle$ be the highest weight vector of the Kac-Moody algebra
$\hat g$ with the weight $j$. Denote also by $V_j$ the Verma module over $\hat
g$
generated by the vector $|j\rangle$. The defining relations for $V_j$ are given
by

$$
\hat n_+|j\rangle =0\,,\qquad H_i|j\rangle =j(H_i)|j\rangle\,,
\eqno{(2.1)}
$$
where $\hat n_+$ is the subalgebra generated by annihilation operators, and
$H_i$
are the Cartan generators.

Let $s_{\alpha}$ be the full set of singular vectors of the module
$V_j$\footnote{The module $V_j$ has singular vectors because the theory is
degenerate by assumption.}. Here $\alpha$ is a parameter labelling these
vectors. They satisfy the following conditions

$$
\hat n_+s_{\alpha}=0\,,\qquad H_is_{\alpha}=j_{\alpha}(H_i)s_{\alpha}\,.
\eqno{(2.2)}
$$

Let the generators of the algebra $\hat g$ be bosonized, i.e. they are
expressed
in terms of free fields. In addition let us assume that one solution of the
equation (2.1) is available via the free fields. Our goal is to find the full
set of the solutions.

Comparing the equation (2.1) and (2.2) we see that the first differs from the
second only due to the parameter $j(H_i)$. By doing the formal transformation
of
the initial weight $j\rightarrow j^{\prime}$ (Weyl reflection) we reduce (2.1)
to (2.2). Thus our problem becomes the one of finding the full set of singular
vectors in the Verma module\footnote{Due to this we denote the solutions of
(2.1) as $\hv{j}{\alpha}$ below. Then the known solution is $\hv{j}{0}$.}.
The last has been solved by Malikov, Feigin and Fuchs [13,14]. It is easy to
see
that the full number of the solutions is $N_s+1$, where $N_s$-the number of the
singular vectors in the Verma module. From physical point of view it means that
the highest weight vector has $N_s+1$-fold degeneracy. It should be noted that
we have to define the solutions of Malikov-Feigin-Fuchs in the bosonized theory
more carefully \footnote{In fact there are expansions of the free fields in
negative (integer) powers.}.

Now let us turn to the construction of conformal blocks. Since there is a one
to
one correspondence between the local fields in the theory and the states in the
Fock space, let $\phi^{(\alpha)}_j$ be the primary field of $G$ WZW model
corresponding to $\hv{j}{\alpha}$. The starting point is the full set of the
representations (solutions) $\{\phi^{(\alpha)}_j\}$. In principle, the
representation $\phi^{(\alpha)}_j$ generates its screening operators
$S^{(\alpha)}$ and identity operator $1^{(\alpha)}$ \footnote{It should be
noted
that there are representations without screening operators and identity
operators. As we shall see in sec.5 the first takes place in the $SL(2)$ WZW
model.}. Following [4,8], one can introduce a BRST-like operator using the
screening operators $S^{(\alpha)}$. Then the irreducible representation
(Hilbert
space of the theory) arises as cohomology groups of this BRST operator.

Let us now construct a general N-point conformal block. First of all introduce
representations for vacua

$$
\vc=1^{(\alpha_1)}\vac\,,\qquad \cvc=\cvac 1^{(\alpha_2)}{}^{\dagger}\,,
\eqno{(2.3)}
$$
where $\vac$ and $\cvac$ are vacua for the free fields, $1^{(\alpha_1)}$ is the
representation of the identity operator and $1^{(\alpha_2)}{}^{\dagger}$ is the
conjugate
identity operator. For a set of physical operators $\{\pf{j_i}{z_i}\}$, one can
choose the set of the representations $\{\rpf{j_i}{\beta_i}{z_i}\}$ (solutions
of eq.(2.1)). In order to take into account the balance of charges (zero modes)
we insert into the correlator a set of the screening operators
$\{S^{(\gamma_i)}\}$ as well as a set of the identity operators
$\{1^{(\lambda_i)}\}$. If this gives the correct balance of charges, then the
conformal block is well defined \footnote{ In fact, it is not hard to see that
we don't define contours for the screening operators in eq.(2.4).  However one
can choose the Felder's contours and take into account the structure of the
corresponding BRST complex.}. Finally, the N-point conformal block is given by

$$
\langle\prod_{i=1}^N\pf{j_i}{z_i}\rangle^{({\vec \alpha},{\vec \beta},{\vec
\gamma},{\vec \lambda})} =\cvac
1^{(\alpha_2)}{}^{\dagger}\prod_{i=1}^N\prod_{m=1}^M\prod_{l=1}^L\rpf{j_i}
{\beta_i}{z_i} S^{(\gamma_m)} 1^{(\lambda_l)}1^{(\alpha_1)}\vac\,.
\eqno{(2.4)}
$$
Here
${\vec \alpha}=(\alpha_1,\alpha_2),{\vec
\beta}=(\beta_1,\beta_2,...\beta_n)$, etc. Notice that above we have
constructed
the full set of basis in the space of N-point conformal blocks.

At present there are really two basis in the space of conformal blocks, namely
the Feigin-Fuchs representation and the Dotsenko-Fateev representation. The
first contains only $\rpf{j}{0}{z}$ (trivial $exp$ of the scalar fields)
representations for physical operators, where as the second has one additional
nontrivial representation $\rpf{j}{1}{z}$. Formally they can be written as

$$
\cvac\prod_{i=1}^N\prod_{m=1}^M\rpf{j_i}{0}{z_i} S^{(0_m)}\vac\,,\qquad
1^{(0)}\equiv 1\,,
\eqno{(2.5)}
$$
and
$$
\cvac 1^{(1)}{}^{\dagger}\rpf{j_1}{1}{z_1}\prod_{i=2}^N\prod_{m=1}^M
\rpf{j_i}{0}{z_i} S^{(0_m)}\vac\,.
\eqno{(2.6)}
$$

As we shall see in sec.5 that more non-trivial representations for
the physical operators are possible and their correlation functions may be
calculated.

\newpage
\section{Wakimoto representation for $SL(2)$ WZW model}

In order to see how our construction works in remainder of this work we shall
focus our attention on the $SL(2)$ WZW model. This is, of course, a
trivial case.  Nevertheless, it contains all technical subtleties which appear
due to bosonization.

As a preparation for a discussion of nontrivial representations in later
sections, let us briefly recall the well-known Wakimoto free field description
of the $SL(2)$ WZW model [5-8,10].

The model is described in terms of one free boson $\varphi$ coupled to a
background charge and a first order bosonic $(\omega,\omega^{\dagger})$ system
of weight (0.1). In terms of mode expansions we have

$$
\vp{z}=x_0+i\a{0}lnz-i\sum_{n=-\infty}^{+\infty\,\,\,\,{}_\prime}\frac{\a{n}}{nz^n}\,,
$$
$$
\om{z}=i\sum_{n=-\infty}^{+\infty}\frac{\o{n}}{z^n}\,,\qquad
\dom{z}=\sum_{n=-\infty}^{+\infty}\frac{\l{n}}{z^{n+1}}\,.
\eqno{(3.1)}
$$

Canonical quantization gives the following commutation relations

$$
[\a{0},x_0]=i\,,\qquad [\a{n},\a{-n}]=n\,,\qquad [\o{n},\l{-n}]=1\,.
\eqno{(3.2)}
$$

Now let us construct the Fock module $\F{j}{0}$.
Define a vector $\hv{j}{0}$ such that

$$
\hv{j}{0}=e^{-2i\alpha_0jx_0}\vac\,,
\eqno{(3.3)}
$$
where the vacuum $\vac$ satisfies the following conditions

$$
\a{n}\vac=\o{n+1}\vac=\l{n}\vac=0\,,\,\,\,n\geq 0\,,
\eqno{(3.4)}
$$
and $2j\in N$.

The Fock space $\F{j}{0}$ is obtained by acting on the vector $\hv{j}{0}$
with the mode $\o{0}$ and all the negative frequency modes of the fields
$\omega,\omega^{\dagger},\varphi$. The basis of $\F{j}{0}$ is given by the
states
$$
\a{-1}^{A_{1}}\a{-2}^{A_{2}}...\o{0}^{B_{0}}\o{-1}^{B_{1}}...
\l{-1}{}^{C_{1}}...\hv{j}{0}\,,
\qquad \{A_i,B_i,C_i\}\in N\,.
\eqno{(3.5)}
$$

Note that there exists a 1-1 correspondence between states in the Fock space
and
fields of the theory. The correspondence is given by

$$
|\phi\rangle=\lim_{z\rightarrow 0}\phi(z)\vac\,.
\eqno{(3.6)}
$$
For example the vector (3.3) corresponds to the field

$$
\phi^j_j(z)=e^{-2i\alpha_0j\vp{z}}\,.
$$
The other important fields are

$$
\phi^j_m(z)=\omega^{j-m}(z) e^{-2i\alpha_0j\vp{z}}\,,\,\,\,\,-j\leq m\leq j\,.
\eqno{(3.6a)}
$$
These fields are the primary fields of the theory.

In the Fock space the $\hat {sl}_2$ algebra is represented by

$$
\e{n}=\l{n}\,,
$$
$$
\n{n}=\frac{1}{2\alpha_0}\a{n}+\sum_{m=-\infty}^{+\infty}:\o{n-m}\l{m}:\,,
\eqno{(3.7)}
$$
$$
\j-{n}=kn\o{n}-\frac{1}{\alpha_0}\sum_{m=-\infty}^{+\infty}\o{n-m}\a{m}-
\sum_{m=-\infty}^{+\infty}\sum_{t=-\infty}^{+\infty}:\o{t}\o{m}\l{n-m-t}:\,.
$$
Here $k$ is the level, $2\alpha_0\,^2=1/k+2$.
It easy to see that the vector $\hv{j}{0}$ is the highest weight vector of the
$\hat {sl}_2$ algebra (3.7) with the weight $j$.

The structure of $\F{j}{0}$ as the module over $\hat {sl}_2$ is shown in fig.1
(see
[7,8]).

\vspace{1cm}
\unitlength=1mm
\special{em:linewidth 0.4pt}
\linethickness{0.4pt}
\begin{picture}(86.00,85.67)
\put(65.33,25.00){\circle*{1.33}}
\put(65.33,45.00){\circle*{1.33}}
\put(65.33,65.00){\circle*{1.33}}
\put(80.33,75.00){\circle*{1.33}}
\put(80.33,85.00){\circle*{1.33}}
\put(80.33,55.00){\circle*{1.33}}
\put(80.33,35.00){\circle*{1.33}}
\put(80.33,36.67){\vector(0,1){16.00}}
\put(80.33,73.00){\vector(0,-1){16.33}}
\put(80.33,77.00){\vector(0,1){6.00}}
\put(65.33,47.00){\vector(0,1){16.00}}
\put(65.33,42.67){\vector(0,-1){15.67}}
\put(79.00,82.67){\vector(-3,-4){11.67}}
\put(78.33,71.67){\vector(-1,-2){11.67}}
\put(78.33,53.00){\vector(-1,-2){12.00}}
\put(78.33,35.67){\vector(-3,2){11.33}}
\put(78.00,55.33){\vector(-4,3){10.67}}
\put(88.00,85.00){\makebox(0,0)[cc]{$|\nu\rangle^{(0)}$}}
\put(86.00,75.00){\makebox(0,0)[cc]{$c_1^{(0)}$}}
\put(87.00,55.00){\makebox(0,0)[cc]{$m_1^{(0)}$}}
\put(86.00,35.00){\makebox(0,0)[cc]{$c_2^{(0)}$}}
\put(60.00,25.00){\makebox(0,0)[cc]{$s_2^{(0)}$}}
\put(60.00,45.00){\makebox(0,0)[cc]{$m_2^{(0)}$}}
\put(60.00,65.00){\makebox(0,0)[cc]{$s_1^{(0)}$}}
\end{picture}
\vspace{-2cm}
\begin{center}
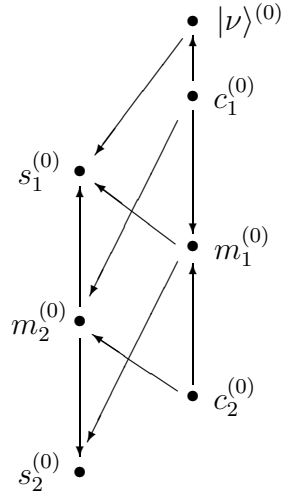
Fig.1: The structure of the Wakimoto module $\F{j}{0}$.
\end{center}
Here and henceforth, arrows go from one vector to another if and only if the
second vector is in the $U_g$ submodule generated by the first one.

It should be noted that the vectors $s_i^{(0)}$ are singular, i.e. they are
annihilated by $\hat n_+$ and have a non-zero weight

$$
\hat n_+s_i^{(0)}=0\,,\qquad J_0^0s_i^{(0)}\neq 0\,.
\eqno{(3.8)}
$$
The vectors $m_i^{(0)}$ and $c_i^{(0)}$ are not singular vectors, $c_i^{(0)}$
are the so-called cosingular vectors. The highest weight vector $\hv{j}{0}
\,\,(\hv{j}{0}\equiv\hv{\nu}{0})$ is both singular and cosingular.

In [7,8] the following complex of the Fock modules was built

$$
{}\stackrel{\q{0}{-2}}{\rightarrow}\f{0}{-1}\stackrel{\q{0}{-1}}{\rightarrow}
\f{0}{0}\stackrel{\q{0}{0}}{\rightarrow}\f{0}{1}\stackrel{\q{0}{1}}{\rightarrow}
\f{0}{2}\stackrel{\q{0}{2}}{\rightarrow}\,,
$$
$$
\q{0}{g+1}\q{0}{g}=0\,,
\eqno{(3.9)}
$$
where
$$
\f{0}{0} \equiv \F{j}{0}\,,\qquad \nu=2j+1\,,\qquad p=k+2\,,
$$
$$
\q{0}{g}=\left\{
\begin{array}{ll}
Q_\nu \,,\quad & g=2n\,,\qquad n\in Z \\
Q_{p-\nu}\,,& g=2n+1\,,
\end{array} \right.
$$
and
$$
\f{0}{g}=\left\{
\begin{array}{ll}
\F{j-pn}{0}\,,\quad & g=2n\,,\qquad n\in Z \\
\F{-j-1-pn}{0}\,,& g=2n+1\,.
\end{array} \right.
$$

The number $g$ is called the ghost number, in analogy to gauge theories. One
can see that $\q{0}{g}$ and $\F{j}{0}$ have the ghost number 1 and 0,
respectively. The BRST charge is defined by

$$
Q_\nu=\prod_{i=1}^{\nu}\oint_{C_i}\,dz_i\dom{z_i}e^{2i\alpha_0\vp{z_i}}\,.
\eqno{(3.10)}
$$
The integration contours for $z_1,...z_\nu$ are shown in fig.2.

\unitlength=1mm
\special{em:linewidth 0.4pt}
\linethickness{0.4pt}
\begin{picture}(95.33,45.00)
\put(101.00,25.00){\circle*{1.33}}
\put(75.00,25.00){\circle*{1.33}}
\bezier{184}(101.00,25.00)(70.67,38.00)(70.00,25.00)
\bezier{180}(101.00,25.00)(72.00,12.00)(70.00,25.00)
\bezier{256}(101.00,25.00)(62.33,45.00)(61.00,25.00)
\bezier{256}(101.00,25.00)(62.33,5.00)(61.00,25.00)
\put(66.00,34.00){\vector(-4,-3){2.00}}
\put(73.33,30.67){\vector(-1,-1){2.67}}
\put(62.00,25.00){\circle*{0.67}}
\put(64.67,25.00){\circle*{0.67}}
\put(67.33,25.00){\circle*{0.67}}
\put(105.33,25.00){\makebox(0,0)[cc]{1}}
\put(80.00,25.00){\makebox(0,0)[cc]{0}}
\put(59.67,32.67){\makebox(0,0)[cc]{$z_1$}}
\put(68.00,29.33){\makebox(0,0)[cc]{$z_{\nu}$}}
\end{picture}
\vspace{-1cm}
\begin{center}
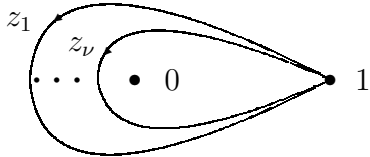
Fig.2: Contours used in the definition of the BRST operator $Q_{\nu}$.
\end{center}

We denote the above complex by ${\bf F}^{(0)}$.

The detail structure of this complex is indicated in fig.3. The horizontal
arrows correspond to the action of the BRST charge.

\unitlength=1.00mm
\special{em:linewidth 0.4pt}
\linethickness{0.4pt}
\begin{picture}(146.33,90.00)
\put(70.00,25.00){\circle*{1.33}}
\put(70.00,45.00){\circle*{1.33}}
\put(70.00,65.00){\circle*{1.33}}
\put(80.00,75.00){\circle*{1.33}}
\put(80.00,85.00){\circle*{1.33}}
\put(80.00,55.00){\circle*{1.33}}
\put(80.00,35.00){\circle*{1.33}}
\put(98.00,45.00){\circle*{1.33}}
\put(98.00,25.00){\circle*{1.33}}
\put(108.00,35.00){\circle*{1.33}}
\put(108.00,55.00){\circle*{1.33}}
\put(108.00,75.00){\circle*{1.33}}
\put(126.00,25.00){\circle*{1.33}}
\put(136.00,35.00){\circle*{1.33}}
\put(136.00,55.00){\circle*{1.33}}
\put(52.00,65.00){\circle*{1.33}}
\put(52.00,55.00){\circle*{1.33}}
\put(52.00,35.00){\circle*{1.33}}
\put(42.00,25.00){\circle*{1.33}}
\put(42.00,45.00){\circle*{1.33}}
\put(24.00,45.00){\circle*{1.33}}
\put(24.00,35.00){\circle*{1.33}}
\put(14.00,25.00){\circle*{1.33}}
\put(5.67,25.00){\vector(1,0){6.67}}
\put(44.00,25.00){\vector(1,0){24.00}}
\put(100.33,25.00){\vector(1,0){23.67}}
\put(138.33,35.00){\vector(1,0){8.00}}
\put(82.33,35.00){\vector(1,0){23.33}}
\put(26.00,35.00){\vector(1,0){23.67}}
\put(26.33,45.00){\vector(1,0){13.67}}
\put(72.67,45.00){\vector(1,0){23.00}}
\put(110.67,55.00){\vector(1,0){23.33}}
\put(54.33,55.00){\vector(1,0){23.00}}
\put(54.33,65.00){\vector(1,0){13.67}}
\put(82.67,75.00){\vector(1,0){23.00}}
\put(80.00,77.33){\vector(0,1){5.67}}
\put(80.00,72.33){\vector(0,-1){15.33}}
\put(80.00,37.33){\vector(0,1){15.67}}
\put(70.00,47.33){\vector(0,1){15.67}}
\put(70.00,42.33){\vector(0,-1){15.33}}
\put(78.00,82.00){\vector(-1,-2){7.33}}
\put(78.33,71.00){\vector(-1,-4){6.00}}
\put(78.33,51.67){\vector(-1,-4){6.00}}
\put(78.00,36.67){\vector(-1,1){6.33}}
\put(78.00,57.00){\vector(-3,4){5.33}}
\put(108.00,57.00){\vector(0,1){15.67}}
\put(108.00,52.33){\vector(0,-1){15.00}}
\put(98.00,27.33){\vector(0,1){15.33}}
\put(106.00,72.00){\vector(-1,-4){6.00}}
\put(106.33,51.33){\vector(-1,-4){5.67}}
\put(106.00,37.00){\vector(-3,4){5.67}}
\put(136.00,37.00){\vector(0,1){15.67}}
\put(134.00,52.00){\vector(-1,-4){6.00}}
\put(52.00,57.00){\vector(0,1){6.00}}
\put(52.00,52.67){\vector(0,-1){15.00}}
\put(42.00,27.00){\vector(0,1){15.67}}
\put(50.33,62.67){\vector(-1,-2){7.67}}
\put(50.00,50.67){\vector(-1,-4){5.67}}
\put(49.67,37.33){\vector(-1,1){6.33}}
\put(24.00,37.00){\vector(0,1){6.00}}
\put(22.00,42.33){\vector(-1,-2){7.67}}
\put(24.00,50.00){\makebox(0,0)[cc]{$\f{0}{-2}$}}
\put(52.00,70.00){\makebox(0,0)[cc]{$\f{0}{-1}$}}
\put(80.00,90.00){\makebox(0,0)[cc]{$\f{0}{0}$}}
\put(108.00,80.00){\makebox(0,0)[cc]{$\f{0}{1}$}}
\put(136.00,60.00){\makebox(0,0)[cc]{$\f{0}{2}$}}
\end{picture}
\vspace{-1cm}
\begin{center}
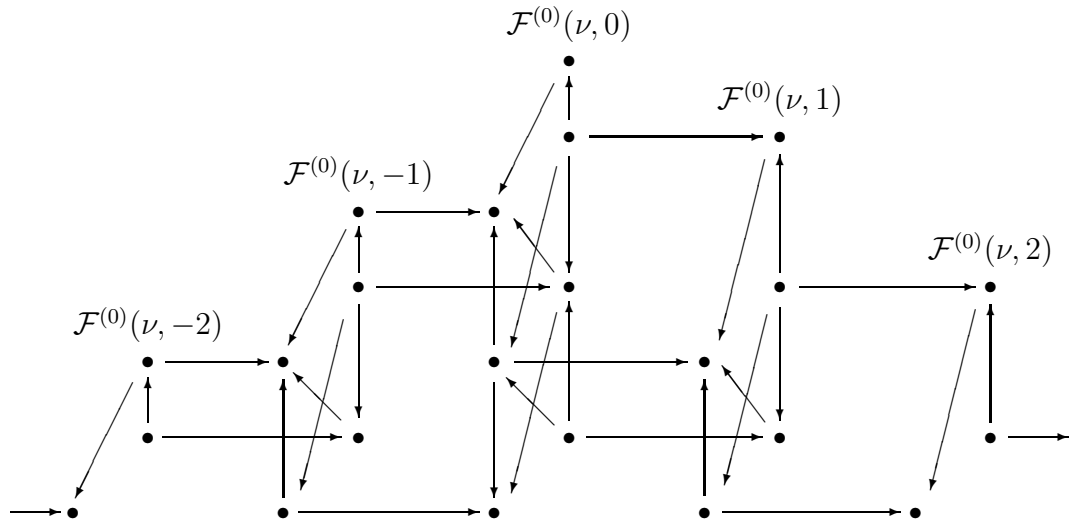
Fig.3: The complex ${\bf F}^{(0)}$. Horizontal arrows indicate where special
vectors are mapped to under the BRST operator. The other arrows indicate the
submodule structure of the Wakimoto modules.
\end{center}

Bernard, Felder and independently Feigin, Frenkel proved that
the cohomology of this complex is entirely located in the middle
Fock space, where it is isomorphic to the irreducible representation
of the $\hat {sl}_2$ algebra. This
irreducible module is the space of physical states of the model.

$$
H^g=\frac{{\rm Ker}\,\q{0}{g}}{{\rm Im}\,\q{0}{g-1}}=\left\{
\begin{array}{ll}
0\,,\qquad & g\neq 0\\
{\cal H}_\nu\,, & g=0\,.
\end{array}\right.
\eqno{(3.11)}
$$
Here ${\cal H}_\nu$ is the Hilbert space or the irreducible highest weight
module.

\section{Dotsenko representation for $SL(2)$ WZW model}

\vspace{0.5cm}

\begin{center}
{\rm 4.1 DOTSENKO MODULES OVER $\hat {sl}_2$ ALGEBRA}
\end{center}

\vspace{0.5cm}

In the previous section we discussed the Wakimoto module over the $\hat {sl}_2$
Kac-Moody algebra and the ${\bf F}^{(0)}$ complex of these modules. We will now
focus on the first nontrivial module over $\hat {sl}_2$, the so-called Dotsenko
module, and on a resolution $\hat {sl}_2$ in terms of Dotsenko modules.

The problem we will address in this section arises quite naturally in the free
field approach to conformal field theories, namely how to construct all
families
of modules over $\hat {sl}_2$ (3.7) in terms of the free fields (3.1) and
complexes whose cohomologies are isomorphic to ${\cal H}$.

A first step in this direction was done by Dotsenko [10]. He found the
conjugate
representation for the highest weight vector using the corresponding Operator
Product Algebra. Explicitly

$$
\tilde{|j\rangle}=\left(\l{-1}\right)^{s+2j}e^{2i\alpha_0(j+s)x_0}\vac\,,
\eqno{(4.1)}
$$
with $s=-k-1$.

Let us say a few words about the Dotsenko representation.
It is clear from the equation (4.1) that the power of $\l{-1}$
is a negative integer\footnote{We consider the case of integrable
representations when $k$ is positive
integer and \\$0\leq j\leq k/2\,,\,\,2j\in N$.}. It means that analytic
continuation
in $s$ is necessary. However the continuation becomes a problem in the case
when
the correlator contains two or more operators in the Dotsenko representation.

It easy to see that the vector (4.1) corresponds to the first singular vector
in the Verma module over $\hat {sl}_2$. Using the technique developed in sec.2
one arrives, after a straightforward calculation, at the following formula for
the first non-trivial representation of the highest weight vector

$$
s_1=\left(\e{-1}\right)^{p-\nu}\hv{j}{0}\stackrel{j\rightarrow
k+1-j}{\rightarrow}\hv{j}{1}=\left(\l{-1}\right)^{2j-k-1}e^{2i\alpha_0(j-k-1)x_0}
\vac\,.
\eqno{(4.2)}
$$
In the above, we also used the formulas (3.3) and (3.7).

Now let us introduce a vector $\hv{j}{1}(s)$ as

$$
\hv{j}{1}(s)=-\frac{1}{2i\pi}\Gamma(2j+1+s)\int_{C}\,dt(-t)^{-2j-1-s}
e^{-t\l{-1}} e^{2i\alpha_0(j-k-1)x_0}\vac\,,
\eqno{(4.3)}
$$
\vspace{1 cm}
where the integration contour is indicated in fig.4.

\unitlength=1mm
\special{em:linewidth 0.4pt}
\linethickness{0.4pt}
\begin{picture}(100.00,35.33)
\put(65.67,25.33){\rule{34.33\unitlength}{0.33\unitlength}}
\bezier{100}(71.67,27.67)(58.33,35.33)(57.00,25.67)
\bezier{92}(71.67,23.00)(57.67,18.00)(57.00,25.67)
\put(100.00,27.67){\vector(-1,0){28.00}}
\put(72.00,23.00){\vector(1,0){27.33}}
\put(71.00,31.67){\makebox(0,0)[cc]{C}}
\put(64.33,24.33){\makebox(0,0)[cc]{0}}
\end{picture}
\vspace{-1.5 cm}
\begin{center}
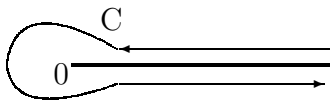
Fig.4: Contour $C$ used in the construction of the Dotsenko representation as
well as the $2$-representation.
\end{center}

In (4.3) $s\in C\,,s\neq  -1,-2,...$ . Our goal is to define the vector (4.3)
as well as a Fock space
(Dotsenko module) in the case of $s=-k-1$.

\newpage
First of all compute the norm of the vector (4.3). Using the formula (A.2) of
the appendix A, one finds

$$
\langle j\hv{j}{1}(s)=\Gamma(2j+1+s)\,.
\eqno{(4.4)}
$$

Let us now define a normalized vector $\hv{\nu}{1}$ as

$$
\hv{\nu}{1}(s)=\hv{j}{1}(s)/ \left( \langle
j\hv{j}{1}(s)\right)^{\frac{1}{2}}\,,\quad \nu=2j+1\,.
\eqno{(4.5)}
$$
The vector $\hv{\nu}{1}(s)$ corresponds to the operator

$$
\rpp{j}{j}{1}{z,s}=-\frac{1}{2i\pi}\Gamma^{1/2}(\nu+s)\int_{C}\,dt(-t)^{-\nu-s}
e^{-t\dom{z}} e^{2i\alpha_0(j-k-1)\vp{z}}\,.
\eqno{(4.6)}
$$
It has the following OP expansions with the currents (3.7)

$$
J^+(z_1)\rpp{j}{j}{1}{z_2,s}=RT\,,
$$
$$
J^0(z_1)\rpp{j}{j}{1}{z_2,s}=\frac{1}{z_{12}}(j+s+k+1)\rpp{j}{j}{1}{z_2,s}+RT\,,
\eqno{(4.7)}
$$
$$
J^-(z_1)\rpp{j}{j}{1}{z_2,s}=\frac{1}{z_{12}^2}\j-{1}\rpp{j}{j}{1}{z_2,s}+
\frac{1}{z_{12}}\j-{0}\rpp{j}{j}{1}{z_2,s}+RT\,,
$$
where
$$
\j-{1}\rpp{j}{j}{1}{z,s}=\frac{s+k+1}{2i\pi}\Gamma^{1/2}(\nu+s)\int_{C}\,dt(-t)^{-2j-s}e^{-t\dom{z}}
e^{2i\alpha_0(j-k-1)\vp{z}}\,,
$$
\begin{eqnarray*}
&&\j-{0}\rpp{j}{j}{1}{z,s}=\frac{1}{\pi}\Gamma^{1/2}(\nu+s)\int_{C}\,dt
(-t)^{-2j-s}e^{-t\dom{z}}e^{2i\alpha_0(j-k-1)\vp{z}}\\
&&\times\left(-\frac{t}{2i}\partial\dom{z}+(j+s+k+1)t^{-1}\om{z}-\frac{1}{2\alpha_0}
\partial\vp{z}\right)\,,
\end{eqnarray*}
$RT$ means terms which are regular as $z_1\rightarrow
z_2\,;z_{12}=z_1-z_2$.

Set

$$
|\j-{1}\rangle (s)=\lim_{z\rightarrow 0}\j-{1}\rpp{j}{j}{1}{z,s}\vac\,,
$$
$$
|\j-{0}\rangle (s)=\lim_{z\rightarrow 0}\j-{0}\rpp{j}{j}{1}{z,s}\vac\,.
$$
Using (A.2) one can easily obtain the norms of these vectors

$$
\langle \j-{1}|\j-{1}\rangle (s)=(2j+s)(k+1+s)^2\,,\quad
\langle \j-{0}|\j-{0}\rangle (s)= 4j^2+(2j+s)(2j+s+k/2)\,.
\eqno{(4.8)}
$$
\newpage
After taking the limit $s\rightarrow  -k-1$, eq.(4.8) becomes

$$
\langle \j-{1}|\j-{1}\rangle =0\,,\quad \langle \j-{0}|\j-{0}\rangle
=8j^2-j(3k+4)+(k+1)(k/2+1)\,.
\eqno{(4.9)}
$$

 From (4.7) and (4.9) we get the following OP relations

$$
J^+(z_1)\rpp{j}{j}{1}{z_2}=RT\,,
$$
$$
J^0(z_1)\rpp{j}{j}{1}{z_2}=\frac{1}{z_{12}}j\rpp{j}{j}{1}{z_2}+RT\,,
\eqno{(4.10)}
$$
$$
J^-(z_1)\rpp{j}{j}{1}{z_2}=\frac{1}{z_{12}}\j-{0}\rpp{j}{j}{1}{z_2}+RT\,,
$$
with $\rpp{j}{j}{1}{z}=\lim_{s\rightarrow -k-1}\rpp{j}{j}{1}{z,s}$.

As can easily be seen from (4.10), $\rpp{j}{j}{1}{z}$ is the primary field
corresponding to the highest weight vector $\hv{\nu}{1}=
\lim_{s\rightarrow -k-1}\hv{\nu}{1}(s)=\lim_{z\rightarrow 0}\rpp{j}{j}{1}{z}
\vac$.

The screening operator in the Dotsenko representation is given by

$$
S^{(1)}=\lim_{s\rightarrow -k-1} S^{(1)}(s)\,,
$$
with
\begin{eqnarray*}
&&S^{(1)}(s)=\oint_{C_z}\,dz:J^+(z)\rpp{-1}{-1}{1}{z,s}:=\\
&&-\frac{1}{2i\pi}(s-1)^{1/2}\Gamma^{1/2}
(s)\oint_{C_z}\,\int_{C}\,dzdt(-t)^{-s}e^{-t\dom{z}}
e^{-2i\alpha_0p\vp{z}}\,.
\end{eqnarray*}

Let $V_j^{(1)}(s)$ be a space generated by the vector
$\hv{\nu}{1}(s)$ and the bosonized generators
\vspace{0.3cm}
(3.7).

{\bf Proposition 4.1}
\vspace{0.3cm}

\hspace{-0.8cm}
{\it Let $v\in V_j^{(1)}(s)$ and $w\in \f{0}{-1}$. Then $\langle w|v\rangle
(s)=0\,,\quad \forall\,v\,,w\,$.}

{\bf Proof}. The proof of this proposition follows from the definitions of the
spaces $V_j^{(1)}(s)$ and $\f{0}{-1}$.

The scalar product of these vectors is given by

$$
\langle w|v\rangle (s)=const\int_{C}\,dt(-t)^{n-s}\,,\qquad n\in Z\,.
$$

Choosing $s$ as $s>0\,,s\in R\,,s>>n$ one can close the integration contour as
shown in fig.5.

\unitlength=1.00mm
\special{em:linewidth 0.4pt}
\linethickness{0.4pt}
\begin{picture}(100.67,40.00)
\put(65.67,25.33){\rule{34.33\unitlength}{0.33\unitlength}}
\bezier{100}(71.67,27.67)(58.33,35.33)(57.00,25.67)
\bezier{92}(71.67,23.00)(57.67,18.00)(57.00,25.67)
\put(100.00,27.67){\vector(-1,0){28.00}}
\put(72.00,23.00){\vector(1,0){27.33}}
\put(71.00,31.67){\makebox(0,0)[cc]{C}}
\put(62.33,23.33){\makebox(0,0)[cc]{0}}
\bezier{56}(100.00,27.67)(100.67,35.33)(97.00,40.00)
\bezier{56}(100.00,23.00)(100.33,15.33)(97.00,10.00)
\put(99.00,36.33){\vector(1,-2){1.00}}
\put(98.67,12.67){\vector(-2,-3){1.33}}
\end{picture}
\begin{center}
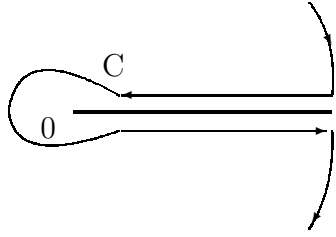
Fig.5: Closing the integration contour $C$ at infinity.
\end{center}

 From this it follows that

$$
\int_{C}\,dt(-t)^{n-s}=\oint_{C}\,dt(-t)^{n-s}=0\,.
$$

Note that the above procedure corresponds to the integration by parts in
the derivation of the OP expansions (4.7). From this point of view, closing
the integration contour seems natural.

Let $\F{j}{1}(s)=V_j^{(1)}(s)\oplus\f{0}{-1}$. Then define the Dotsenko module
as

$$
\F{j}{1}=\lim_{s\rightarrow -k-1}\F{j}{1}(s)\,.
$$

Now let us turn to the structure of the Dotsenko module as a $\hat {sl}_2$
module.  Consider the following set of vectors

$$
s_1^{(1)}=\left(\e{-1}\right)^{p-\nu}\hv{\nu}{1}(s)\,,\quad
s_2^{(1)}=\left(\j-{0}\right)^{\nu} \hv{\nu}{1}(s)\,, $$
$$
s_3^{(1)}=\left(\e{-1}\right)^{p+\nu}\left(\j-{0}\right)^{\nu}\hv{\nu}{1}(s)\,,\quad
s_4^{(1)}=\left(\j-{0}\right)^{2p-\nu}\left(\e{-1}\right)^{p-\nu}\hv{\nu}{1}(s)\,,
$$
$$
s_5^{(1)}=...
$$

It is straightforward to see that, under $s\rightarrow -k-1$ $\{s_i^{(1)}\}
\rightarrow$ the singular vectors of the Verma module.

In terms of the free fields we have the following expressions for $s_1^{(1)}$
and
$s_2^{(1)}$

$$
s_1^{(1)}=\frac{(-)^{p-2j}}{2i\pi}
(\nu+s)^{1/2}...(p+s-1)^{1/2}\Gamma^{1/2}(p+s)
\int_{C}\,dt(-t)^{-p-s}e^{-t\l{-1}}e^{2i\alpha_0(j+s)x_0}\vac\,,
$$
$$
s_2^{(1)}=-\frac{1}{2i\pi}\Gamma^{1/2}(1+s)\int_{C}\,dt(-t)^{-1-s}
\left(\frac{1}{\alpha_0}t^{-1}\a{-1}-t^2\l{-2}\right) e^{-t\l{-1}}
e^{2i\alpha_0(j+s)x_0}\vac\,.
$$

Their norms are given by
\newpage
$$
\langle s_1|s_1\rangle ^{(1)} (s)=\Gamma (s+p)/\Gamma (\nu+s)\,,\qquad
\langle s_2|s_2\rangle ^{(1)} (s)=s(2p+s-1)\,.
\eqno{(4.11)}
$$

In the above, we consider the $j=0$ case for the vector $s_2^{(1)}$ because its
explicit expression in terms or the free fields has a rather complicated form.
However the generalization to a general $j$ is straightforward.  After taking
the limit $s\rightarrow -k-1$, eq.(4.11) becomes

$$
\langle s_1|s_1\rangle ^{(1)}=0\,,\qquad \langle s_2|s_2\rangle ^{(1)}\neq 0\,.
\eqno{(4.12)}
$$

In an analogous way, one can calculate the norms of $s_3$ and $s_4$
$$
\langle s_3|s_3\rangle ^{(1)}= \langle s_4|s_4\rangle ^{(1)}=0\,.
\eqno{(4.13)}
$$

 From (4.12) and (4.13) it follows that the structure of $V_j^{(1)}$ is
described by the
\vspace{0.8 cm}
diagram shown in fig.6.

\unitlength=1mm
\special{em:linewidth 0.4pt}
\linethickness{0.4pt}
\begin{picture}(85.00,35.67)
\put(80.33,19.67){\circle*{1.33}}
\put(80.33,35.00){\circle*{1.33}}
\put(80.33,32.67){\vector(0,-1){11.00}}
\put(89.00,35.00){\makebox(0,0)[cc]{$|\nu\rangle^{(1)}$}}
\put(89.00,19.67){\makebox(0,0)[cc]{$s_2^{(1)}$}}
\end{picture}
\vspace{-1.5cm}
\begin{center}
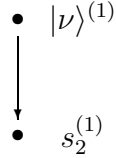
Fig.6: The submodule structure of $V_j^{(1)}$.
\end{center}

Finally, the structure of the Dotsenko module is given by the diagram presented
in fig.7.

\vspace{-0.5cm}
\unitlength=1mm
\special{em:linewidth 0.4pt}
\linethickness{0.4pt}
\begin{picture}(85.00,-5.67)
\put(65.00,-85.00){\circle*{1.33}}
\put(65.00,-65.00){\circle*{1.33}}
\put(65.00,-45.00){\circle*{1.33}}
\put(80.00,-35.00){\circle*{1.33}}
\put(80.00,-25.00){\circle*{0.00}}
\put(80.67,-25.00){\circle*{0.00}}
\put(80.00,-25.00){\circle*{1.33}}
\put(80.00,-15.00){\circle*{1.33}}
\put(80.00,-5.00){\circle*{1.33}}
\put(80.00,-55.00){\circle*{1.33}}
\put(80.00,-75.00){\circle*{1.33}}
\put(80.00,-74.67){\vector(0,1){16.00}}
\put(80.00,-38.67){\vector(0,-1){15.00}}
\put(80.00,-33.00){\vector(0,1){6.00}}
\put(80.00,-7.00){\vector(0,-1){6.33}}
\put(65.00,-63.00){\vector(0,1){16.00}}
\put(65.00,-68.67){\vector(0,-1){15.33}}
\put(77.33,-28.00){\vector(-2,-3){9.67}}
\put(78.00,-38.33){\vector(-1,-2){12.00}}
\put(78.00,-58.00){\vector(-1,-2){11.00}}
\put(77.67,-74.00){\vector(-3,2){10.67}}
\put(77.67,-54.00){\vector(-4,3){10.33}}
\put(87.00,-5.00){\makebox(0,0)[cc]{$|\nu\rangle^{(1)}$}}
\put(86.00,-15.00){\makebox(0,0)[cc]{$s_2^{(1)}$}}
\put(93.00,-25.00){\makebox(0,0)[cc]{$|2p-\nu\rangle^{(0)}$}}
\put(86.00,-35.00){\makebox(0,0)[cc]{$c_1^{(0)}$}}
\put(86.00,-55.00){\makebox(0,0)[cc]{$m_1^{(0)}$}}
\put(86.00,-75.00){\makebox(0,0)[cc]{$c_2^{(0)}$}}
\put(60.00,-85.00){\makebox(0,0)[cc]{$s_2^{(0)}$}}
\put(60.00,-65.00){\makebox(0,0)[cc]{$m_2^{(0)}$}}
\put(60.00,-45.00){\makebox(0,0)[cc]{$s_1^{(0)}$}}
\put(32.00,-15.00){\makebox(0,0)[cc]{Fig.7: The submodule structure of}}
\put(38.00,-20.00){\makebox(0,0)[cc]{the Dotsenko module $\F{j}{1}$.}}
\end{picture}

\vspace{10cm}

An important remark is that the Dotsenko module contains the Verma module part
at the top and the Wakimoto module part at the bottom.

\vspace{0.5cm}

\begin{center}
{\rm 4.2 BRST-LIKE COMPLEX OF DOTSENKO MODULES }
\end{center}

\vspace{0.5cm}

Let us now look at a BRST-like complex of the Dotsenko modules.

We first define the modules with the negative ghost numbers. They have the same
structure as shown in fig.7.
Set

$$
\f{1}{q}=\left\{
\begin{array}{ll}
\F{j-pn}{1}\,,\quad & q=-2n\,,\qquad n\in Z_+ \\
\F{-j-1-pn}{1}\,,& q=-2n-1\,,
\end{array} \right.
\eqno{(4.14)}
$$
where $\f{1}{0}\equiv\F{j}{1}$. The number $q$ is called the ghost number, in
analogy to gauge theories.

Now let us turn to the construction of the Dotsenko modules with the positive
ghost numbers. The starting point is the Wakimoto modules $\f{0}{q-1}$ with
$q>0$. Let $\{ c_i\,,i>1\}$ be a subset of cosingular vectors of the Wakimoto
module $\f{0}{q-1}$(see fig.1). These cosingular vectors generate a space
$\f{1}{q}$. By construction of $\f{1}{q}$
\vspace{1cm}
one has the arrows of the diagram shown in fig.8.

\unitlength=1mm
\special{em:linewidth 0.4pt}
\linethickness{0.4pt}
\begin{picture}(92.67,81.00)
\put(70.00,20.00){\circle*{1.33}}
\put(70.00,40.00){\circle*{1.33}}
\put(70.00,60.00){\circle*{1.33}}
\put(70.00,80.33){\circle*{1.33}}
\put(85.00,70.00){\circle*{1.33}}
\put(85.00,50.00){\circle*{1.33}}
\put(85.00,30.00){\circle*{1.33}}
\put(85.00,48.00){\vector(0,-1){16.00}}
\put(85.00,52.00){\vector(0,1){15.67}}
\put(85.00,67.67){\vector(0,0){0.00}}
\put(83.00,71.33){\vector(-3,2){11.00}}
\put(70.00,62.33){\vector(0,1){15.67}}
\put(70.00,57.67){\vector(0,-1){15.67}}
\put(70.00,22.00){\vector(0,1){15.33}}
\put(82.67,31.67){\vector(-3,2){10.67}}
\put(83.00,51.33){\vector(-3,2){11.00}}
\put(83.00,68.00){\vector(-1,-2){12.33}}
\put(83.00,47.33){\vector(-1,-2){12.00}}
\put(92.67,30.00){\makebox(0,0)[cc]{$m_3^{(0)}$}}
\put(92.67,50.00){\makebox(0,0)[cc]{$c_2^{(0)}$}}
\put(92.67,70.00){\makebox(0,0)[cc]{$m_1^{(0)}$}}
\put(63.67,80.33){\makebox(0,0)[cc]{$s_1^{(0)}$}}
\put(63.67,60.00){\makebox(0,0)[cc]{$m_2^{(0)}$}}
\put(63.67,40.00){\makebox(0,0)[cc]{$s_2^{(0)}$}}
\put(63.67,20.00){\makebox(0,0)[cc]{$m_4^{(0)}$}}
\end{picture}
\vspace{-1 cm}
\begin{center}
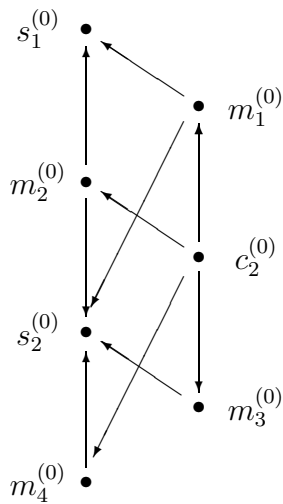
Fig.8: The submodule structure of the Dotsenko modules $\f{1}{q}\,\,,\,q>0$.
\end{center}
\newpage
{\bf Proposition 4.2}
\vspace{0.3cm}

\hspace{-0.8cm}
{\it The following infinite sequence}

$$
{}\stackrel{\q{1}{-2}}{\rightarrow}\f{1}{-1}\stackrel{\q{1}{-1}}{\rightarrow}
\f{1}{0}\stackrel{\q{1}{0}}{\rightarrow}\f{1}{1}\stackrel{\q{1}{1}}{\rightarrow}
\f{1}{2}\stackrel{\q{1}{2}}{\rightarrow}\,,
$$
$$
\q{1}{q+1}\q{1}{q}=0\,,
$$
$$
\q{1}{q}=\left\{
\begin{array}{ll}
Q_{p-\nu} \,,\quad & q=2n\,,\qquad n\in Z \\
Q_{\nu}\,,& q=2n+1\,,
\end{array} \right.
$$
{\it is a complex.}

We denote this complex by ${\bf F}^{(1)}$. Notice that one has a relation
between the ghost numbers of ${\bf F}^{(0)}$ and ${\bf F}^{(1)}$ complexes.

$$
g=q-1\,.
\eqno{(4.15)}
$$

{\bf Proof}. The proof of this proposition follows from the structure
of the ${\bf F}^{(0)}$ complex, our construction and an additional
result presented below. The detailed structure \vspace{1.5cm}
of ${\bf F}^{(1)}$ can be pictured as shown in fig.9.

\unitlength=1mm
\special{em:linewidth 0.4pt}
\linethickness{0.4pt}
\begin{picture}(145.33,85.00)
\put(70.00,20.00){\circle*{1.33}}
\put(70.00,40.00){\circle*{1.33}}
\put(80.00,30.00){\circle*{1.33}}
\put(80.00,50.00){\circle*{1.33}}
\put(80.00,60.00){\circle*{1.33}}
\put(80.00,70.00){\circle*{1.33}}
\put(80.00,80.00){\circle*{1.33}}
\put(98.00,60.00){\circle*{1.33}}
\put(98.00,40.00){\circle*{1.33}}
\put(98.00,20.00){\circle*{1.33}}
\put(108.00,30.00){\circle*{1.33}}
\put(108.00,50.00){\circle*{1.33}}
\put(126.00,40.00){\circle*{1.33}}
\put(126.00,20.00){\circle*{1.33}}
\put(136.00,30.00){\circle*{1.33}}
\put(52.00,30.00){\circle*{1.33}}
\put(52.00,40.00){\circle*{1.33}}
\put(52.00,50.00){\circle*{1.33}}
\put(52.00,70.00){\circle*{1.33}}
\put(42.00,20.00){\circle*{1.33}}
\put(24.00,20.00){\circle*{1.33}}
\put(24.00,30.00){\circle*{1.33}}
\put(24.00,50.00){\circle*{1.33}}
\put(26.67,20.00){\vector(1,0){13.33}}
\put(72.33,20.00){\vector(1,0){23.33}}
\put(128.67,20.00){\vector(1,0){16.67}}
\put(110.33,30.00){\vector(1,0){23.67}}
\put(54.33,30.00){\vector(1,0){23.33}}
\put(13.33,30.00){\vector(1,0){8.33}}
\put(54.33,40.00){\vector(1,0){14.00}}
\put(100.33,40.00){\vector(1,0){23.67}}
\put(26.33,50.00){\vector(1,0){23.00}}
\put(82.67,50.00){\vector(1,0){23.00}}
\put(82.67,60.00){\vector(1,0){13.33}}
\put(54.33,70.00){\vector(1,0){23.67}}
\put(80.00,78.00){\vector(0,-1){6.33}}
\put(80.00,51.67){\vector(0,1){6.33}}
\put(80.00,47.67){\vector(0,-1){16.00}}
\put(70.00,22.00){\vector(0,1){16.00}}
\put(78.33,57.33){\vector(-1,-2){7.67}}
\put(78.33,46.67){\vector(-1,-4){6.00}}
\put(77.67,32.67){\vector(-1,1){5.67}}
\put(52.00,68.00){\vector(0,-1){16.00}}
\put(52.00,31.67){\vector(0,1){6.33}}
\put(50.33,37.33){\vector(-1,-2){7.67}}
\put(24.00,47.33){\vector(0,-1){15.33}}
\put(98.00,41.67){\vector(0,1){16.00}}
\put(98.00,38.00){\vector(0,-1){16.00}}
\put(107.00,51.67){\vector(-1,1){7.33}}
\put(106.33,31.00){\vector(-1,1){7.33}}
\put(126.00,22.67){\vector(0,1){15.33}}
\put(108.00,32.00){\vector(0,1){15.67}}
\put(106.33,46.33){\vector(-1,-4){6.00}}
\put(133.67,31.67){\vector(-1,1){5.67}}
\put(66.67,67.00){\makebox(0,0)[cc]{(-1)}}
\put(38.33,47.00){\makebox(0,0)[cc]{(-2)}}
\put(15.33,27.00){\makebox(0,0)[cc]{(-3)}}
\put(24.00,55.00){\makebox(0,0)[cc]{$\f{1}{-2}$}}
\put(52.00,75.00){\makebox(0,0)[cc]{$\f{1}{-1}$}}
\put(80.00,85.00){\makebox(0,0)[cc]{$\f{1}{0}$}}
\put(98.00,65.00){\makebox(0,0)[cc]{$\f{1}{1}$}}
\put(126.00,45.00){\makebox(0,0)[cc]{$\f{1}{2}$}}
\end{picture}
\vspace{-1cm}
\begin{center}
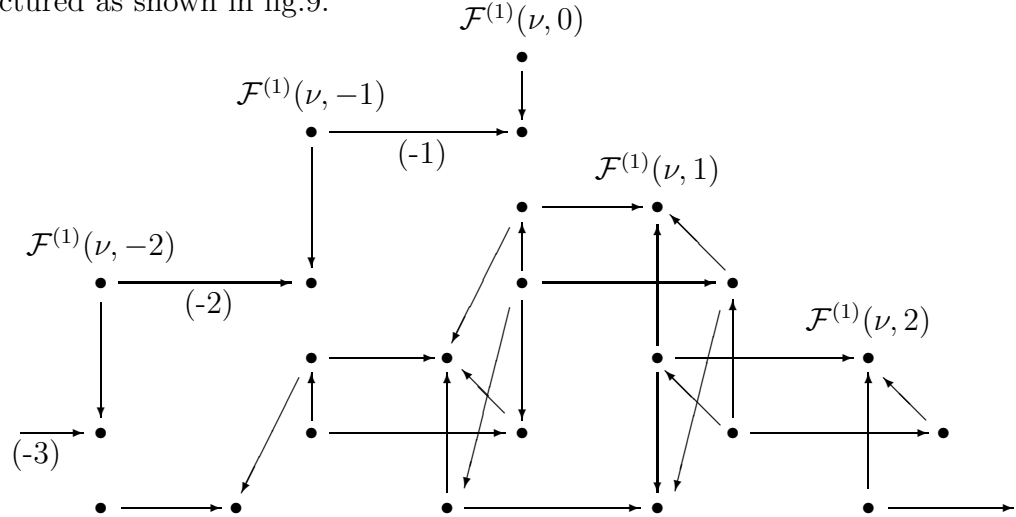
Fig.9: The complex ${\bf F}^{(1)}$. Horizontal arrows indicate where special
vectors are mapped to under the BRST operator. The other arrows indicate
the submodule structure of the Dotsenko modules.
\end{center}

By construction of ${\bf F}^{(1)}$, the horizontal arrows of its bottom part
are ones of the ${\bf F}^{(0)}$ complex (see fig.3).
\newpage
{\bf Proposition 4.3}
\vspace{0.3cm}

\hspace{-0.8cm}
{\it There are a sequence of horizontal arrows} $(-i)$.

\vspace{0.3cm}
{\bf Proof}. Let us consider the arrow $(-1)$. In this case it is rewritten as
(see fig.9)

$$
|v\rangle\sim Q_{\nu}\hv{-\nu}{1}\,,\qquad
|v\rangle=\left(\j-{0}\right)^{\nu}\hv{\nu}{1}\,.
$$
For simplicity we set $j=0\,(\nu=1)$.

Next define vectors

$$
|w\rangle (s)=-\frac{1}{2i\pi}\Gamma^{1/2}(s-1)
\int_{C}\,dt(-t)^{1-s}e^{-t\l{-1}} e^{2i\alpha_0px_0}\vac\,,
$$
$$ |v\rangle (s)=-\frac{1}{2i\pi}\Gamma^{1/2}(s+1)
\int_{C}\,dt(-t)^{-s}\left(-\frac{1}{\alpha_0}\a{-1}+t\l{-2}\right)e^{-t\l{-1}}
e^{2i\alpha_0(p-1)x_0}\vac\,.
$$
It is easy to see that $|w\rangle (s)\rightarrow \hv{-1}{1}\,,|v\rangle
(s)\rightarrow \j-{0}\hv{1}{1}$ under $s\rightarrow -k-1$.

Compute the matrix element $\langle v|\q{1}{-1}|w\rangle (s)$, here the BRST
operator $\q{1}{-1}=Q_1$ is defined in (3.10).  Using the integral (A.2) one
gets

$$
\langle v|Q_1|w\rangle (s)=(s(s-1))^{1/2}\,.
\eqno{(4.16)}
$$
After taking the limit $s\rightarrow -k-1$, eq.(4.16) becomes

$$
\langle 1|\j-{0},\,Q_1\hv{-1}{1} =(p(p-1))^{1/2}\,.
\eqno{(4.17)}
$$
The same calculation can be repeated for a general $j$ as well as $i$.

Finally, one can compute the cohomology of this complex

$$
H^q=\frac{{\rm Ker}\,\q{1}{q}}{{\rm Im}\,\q{1}{q-1}}=\left\{
\begin{array}{ll}
0\,,\qquad & q\neq 0\\
{\cal H}_\nu\,, & q=0\,.
\end{array}\right.
\eqno{(4.18)}
$$

\hspace{-0.8cm}
where ${\cal H}_\nu$ is the irreducible module of $\hat {sl}_2$ (Hilbert space
of the model).

\vspace{0.5cm}

\begin{center}
{\rm 4.3 CONFORMAL BLOCKS ON THE PLANE}
\end{center}

\vspace{0.5cm}

In this section we use the Dotsenko representation together with the Wakimoto
representation to give an explicit calculations for conformal blocks on the
plane.

As an example we compute the two-point function of the primary fields. This is
a
trivial case, but it contains all technical subtleties which appear due to
using
the Dotsenko representation.

Set $\vc =\vac \,, \cvc =\cvac 1^{(1)}{}^{\dagger}$, where $1^{(1)}$ is
the Dotsenko representation for the identity operator. The simplest two-point
function is given by

$$
\langle \phi^j_{-j}(z_1)\phi^j_j(z_2)\rangle \,.
$$

Choose the following representation for the primary fields:
$\rpp{-j}{j}{0}{z_1}$ and $\rpp{j}{j}{1}{z_2}$. It is easy to see that one does
not need additional screening operators as well as identity operators. Finally,
the two-point function is written as

$$
\langle\phi ^j_{-j}(z_1)\phi ^j_j(z_2)\rangle^{({\vec\alpha},{\vec\beta},
{\vec\gamma},{\vec\lambda})}=\cvac 1^{(1)}{}^{\dagger} \rpp{-j}{j}{0}{z_1}
\rpp{j}{j}{1}{z_2} \vac\,.
\eqno{(4.19)}
$$
Here ${\vec\alpha}=(0,1)\,,{\vec\beta}=(0,1)\,,{\vec\gamma}={\vec\lambda}= 0$.

Then using the following two-point functions

$$
\langle \om{z_1}\dom{z_2}\rangle=i/z_{12}\,,\quad \langle
\vp{z_1}\vp{z_2}\rangle =-\ln (z_{12})\,,
\eqno{(4.20)}
$$
as well as the explicit formulas (3.6a), (4.6) for the primary fields and
taking
the limit $s\rightarrow -k-1$, one finds

$$
\langle\phi ^j_{-j}(z_1)\phi ^j_j(z_2)\rangle^{({\vec\alpha},{\vec\beta},
{\vec\gamma},{\vec\lambda})}=C_1/z_{12}^{2\triangle}\,,\qquad
C_1=i^{4j+1}(k...(k-2j))^{\frac{1}{2}}
\eqno{(4.21)}
$$
Here $\triangle =j(j+1)/k+2$.
$C_1$ can be absorbed into a proper normalization of the primary fields.

In order to complete the representation of the conformal blocks on the plane
let
us in the remaining part of this section sketch the BRST invariant chiral
primary fields for the Dotsenko representation.

Following [4,8], we introduce a chiral primary field $
{}_m\Phi_{\nu\rho}^{\mu}{}^{(1)}$ as

$$
{}_m\Phi _{\nu\rho}^{\mu}{}^{(1)}\,:\, {\cal H}_\nu\rightarrow  {\cal H}_\mu\,,
\eqno{(4.22)}
$$
where $\nu=2j+1\,,\rho=2j_1+1\,,\mu=2j_2+1$.

In terms of the free fields one has

$$
{}_m\Phi _{\nu\rho}^{\mu}{}^{(1)}(z)=\rpp{m}{j}{1}{z}\prod_{i=1}^{r}\oint_{C_i}
\,dz_i\dom{z_i}e^{2i\alpha_0\vp{z_i}}\,,
\eqno{(4.23)}
$$
where the field $\rpp{m}{j}{1}{z}$ corresponds to the state
$(\j-{0})^{j-m}\hv{\nu}{1}$ of the Dotsenko module. A number of screening
operators $r$ is determined by the balance of charges in a three-point function
(matrix element of the operator (4.23))\footnote{Note that $r$ depends on a
representation for the three-point conformal block, namely $r({\vec\beta},
{\vec\gamma},{\vec\lambda})$}. $C_i$ are the integration contours of the
Felder's type.

Following the method used by Bernard and Felder in [4,8], one can prove that
the
chiral field (4.23) commutes with the BRST operator (3.10). In particular, one
finds

$$
Q_\mu {}_m\Phi _{\nu\rho}^{\mu}{}^{(1)}(z)=e^{\frac{i\pi\mu(\nu+1)}{k+2}}
{}_m\Phi _{\nu-\rho}^{-\mu}{}^{(1)}(z)Q_\rho\,.
\eqno{(4.24)}
$$

It should be noted that a phase factor in (4.24) is due to the free scalar
field
only. This is so, because for $s=-k-1$ the $\omega\,,\omega^{\dagger}$ fields
do not give rise to nontrivial monodromy. This completes the representation
of the conformal blocks on the plane.

\section{One more representation for $SL(2)$
WZW model}

\vspace{0.5cm}

\begin{center}
{\rm 5.1 NEW MODULES OVER $\hat {sl}_2$ ALGEBRA}
\end{center}

\vspace{0.5cm}

In this section we want to derive another nontrivial representation for the
$SL(2)$ WZW model.

Our starting point is the second singular vector of the Verma
module\footnote{We
shall call the representation $\phi^{(\alpha)}_j$ as the
$\alpha$-representation here and below. Then the Wakimoto representation is the
$0$-representation, the Dotsenko  - the $1$-representation, etc.}. Explicitly

$$
s_2=(\j-{0})^{\nu}\hv{j}{0}\,.
\eqno{(5.1)}
$$

After a straightforward calculation one arrives at the following formula for
the
second nontrivial representation of the highest weight vector

$$
\hv{j}{2}=(\j-{0})^{-2j-1}\hv{-j-1}{0}\,.
\eqno{(5.2)}
$$

Following the method used in sec.4 for the Dotsenko representation, one can
define a vector $\hv{j}{2}(s)$ as

$$
\hv{j}{2}(s)=-\frac{1}{2i\pi}
\Gamma(1+s-2j)\int_{C}\,dt(-t)^{2j-s-1}(1+t\o{0})^{-2j-2}e^{2i\alpha_0(j+1)x_0}\vac
\,,
\eqno{(5.3)}
$$
where the integration contour C is shown in fig.4, and where $s$ is an
arbitrary noninteger parameter. Also, we used the formulas (3.5) and (3.7) in
the above.

Note that, in comparison with the Dotsenko representation, the
$2$-representation has simpler expressions for the primary fields and the
singular vector. Namely

$$
\hv{j,m}{2}(s)=-\frac{1}{2i\pi}\Gamma(1+s-j-m)\int_{C}\,dt(-t)^{j+m-s-1}
(1+t\o{0})^{-2j-2}e^{2i\alpha_0(j+1)x_0}\vac\,,
\eqno{(5.4)}
$$
$$
s_2=-\frac{1}{2i\pi}\Gamma(2+s)\int_{C}\,dt(-t)^{-s-2}
(1+t\o{0})^{-2j-2}e^{2i\alpha_0(j+1)x_0}\vac\,.
\eqno{(5.5)}
$$
Here $\hv{j}{2}(s)\equiv\hv{j,j}{2}(s)$.

\newpage
The normalized vector $\hv{\nu}{2}(s)=\hv{j}{2}/(\langle j\hv{j}{2}(s))^{1/2}$
corresponds to the operator

$$
\rpp{j}{j}{2}{z,s}=-\frac{1}{2i\pi}N_j\int_{C}\,dt(-t)^{2j-s-1}(1-it\om{z})^{-2j-2}
e^{2i\alpha_0(j+1)\vp{z}}\,,
\eqno{(5.6)}
$$
where $N_j=\Gamma^{1/2}(1+s-2j)\Gamma(2+2j)/\Gamma(s+3)$.

It has the following OP expansions with the currents (3.7)

$$
J^+(z_1)\rpp{j}{j}{2}{z_2,s}=\frac{(1+s)N_j}{2i\pi z_{12}}
\int_{C}\,dt(-t)^{2j-s}(1-it\om{z_2})^{-2j-2}
e^{2i\alpha_0(j+1)\vp{z_2}}+RT\,, $$
$$
J^0(z_1)\rpp{j}{j}{2}{z_2,s}=-\frac{(j-s-1)N_j}{2i\pi z_{12}}
\int_{C}\,dt(-t)^{2j-s-1}(1-it\om{z_2})^{-2j-2}e^{2i\alpha_0(j+1)\vp{z_2}}+RT\,,
\eqno{(5.7)}
$$
$$
J^-(z_1)\rpp{j}{j}{2}{z_2,s}=-\frac{(1+s-2j)N_j}{2i\pi z_{12}}
\int_{C}\,dt(-t)^{2j-s-2}(1-it\om{z_2})^{-2j-2}e^{2i\alpha_0(j+1)\vp{z_2}}+RT\,.
$$
In terms of states the equation (5.7) is given by

$$
J^{\alpha}_n\hv{\nu}{2}(s)=0\,,\qquad \alpha =\{+,-,0\}\,,\qquad n>0\,.
\eqno{(5.7a)}
$$

Set

$$
|i\rangle (s)=\left\{
\begin{array}{ll}
\left(\e{0}\right)^i\hv{\nu}{2}(s)\,,\qquad i>0\,,\qquad i\in N\,, \\
\hv{\nu}{2}(s)\,,\qquad i=0\,, \\
\left(\j-{0}\right)^{|i|}\hv{\nu}{2}(s)\,,\qquad i<0\,.
\end{array} \right.
\eqno{(5.8)}
$$
Using (3.7) and (B.4) one can easily get the norms of these vectors

$$
\langle i|i\rangle (s)=\left\{
\begin{array}{ll}
(s-2j)...(s-2j-i+1)\,,\qquad i>0\,, \\
1\,,\qquad i=0\,, \\
(s+1-2j)...(s-i-2j)/(s+2)^2...(s+1-i)^2 \,,\qquad i<0\,.
\end{array} \right.
\eqno{(5.9)}
$$
The connections between these vectors can be illustrated by the following
diagram

$$
\cdot\cdot\cdot\longleftrightarrow|2\rangle (s)
\longleftrightarrow|1\rangle (s)
\longleftrightarrow|0\rangle (s)
\longleftrightarrow|-1\rangle (s)
\longleftrightarrow|-2\rangle (s)\longleftrightarrow\cdot\cdot\cdot
$$
Here the arrows denote an action of $\e{0}$ and $\j-{0}$.

After taking the limit $s\rightarrow -1$, eq.(5.9) reduces to

$$
\langle i|i\rangle =\left\{
\begin{array}{ll}
(-)^i(2j+1)...(2j+i)\,,\qquad i>0\,, \\
1\,,\qquad i=0\,, \\
(-)^i 2j...(2j+1+i)/(i!)^2 \,,\qquad -2j\le i<0\,.
\end{array} \right.
\eqno{(5.10)}
$$

In fact, it is not hard to see that the previous diagram becomes

$$
\cdot\cdot\cdot\longleftrightarrow|2\rangle
\longleftrightarrow|1\rangle
\longleftarrow|0\rangle
\longleftrightarrow|-1\rangle
\longleftrightarrow|-2\rangle
\longleftrightarrow\cdot\cdot\cdot
\longleftrightarrow|-2j\rangle
$$

As a consequence of the eq.(5.10) the vector $\hv{\nu}{2}=\lim_{s\rightarrow
-1}\hv{\nu}{2}(s)$ is not the highest weight vector. One has

$$
J^{\alpha}_n\hv{j}{2}=0\,,\qquad \alpha =\{+,-,0\}\,,\qquad n>0\,,
$$
$$
\e{0}\hv{\nu}{2}\neq 0\,,\qquad \n{0}\hv{\nu}{2}=j\hv{\nu}{2}\,,
\qquad\j-{0}\hv{\nu}{2}\neq 0\,.
\eqno{(5.11)}
$$

Let $\hat n_-^{\prime}$ be a subalgebra on the creation operators of $\hat
{sl}_2$ without $\j-{0}$. Let $V_i(s)$ be a space generated by the vector
$|i\rangle (s)$ and the bosonized generators of $\hat n_-^{\prime}$. Define a
space $V_j(s)$ as $V_j(s)=\sum_{i=-\infty}^{+\infty}\oplus V_i(s)$. The algebra
$sl_2$ acts on $ V_i(s)$ as follows

$$
\e{0}\,:V_i(s)\rightarrow V_{i+1}(s)\oplus V_i \,,\qquad
\n{0}\,:V_i(s)\rightarrow V_i \,,\qquad
\j-{0}\,:V_i(s)\rightarrow V_{i-1}(s)\oplus V_i \,.
$$

It is not hard to prove that in the case of $s=-1$ the space
$V=\lim_{s\rightarrow -1}V_j(s)$ has the same structure as given in the above
diagram.

Now one can define a space $V_j^{(2)}$ as $V_j^{(2)}=V/SV$, where $SV=
\sum_{i=1}^{+\infty}\oplus V_i\,,\qquad V_i=\lim_{s\rightarrow -1}V_i(s)$.
As a consequence of the above, the vector $\hv{\nu}{2}$ becomes the highest
weight vector (with respect to $V_j^{(2)}$).

We now proceed in complete accordance with the construction of the Dotsenko
module.

{\bf Proposition 5.1}

\hspace{-0.8cm}
{\it Let $v\in V_j^{(2)}(s)$ and $w\in \f{0}{1}$. Then $\langle w|v\rangle
=0\,,\quad \forall v\,,w\,$.\\}
\vspace{0.2cm}

{\bf Proof}. The proof is the same as one of the Proposition 4.1.

Define the $2$-module as

$$
\F{j}{2}=V_j^{(2)}\oplus\f{0}{1}\,.
$$

Now let us turn to the structure of this module as a $\hat {sl}_2$ module.

Consider the following set of vectors

$$
s_1^{(2)}=\left(\e{-1}\right)^{p-\nu}\hv{\nu}{2}(s)\,,\quad
s_2^{(2)}=\left(\j-{0}\right)^{\nu} \hv{\nu}{2}(s)\,, $$
$$
s_3^{(2)}=\left(\e{-1}\right)^{p+\nu}\left(\j-{0}\right)^{\nu}\hv{\nu}{2}(s)\,,\quad
s_4^{(2)}=\left(\j-{0}\right)^{2p-\nu}\left(\e{-1}\right)^{p-\nu}\hv{\nu}{2}(s)\,,
$$
$$
s_5^{(2)}=...
$$

It is straightforward to see that, under $s\rightarrow -1$ $\{ s_i^{(2)}\}
\rightarrow$ the singular vectors of the Verma module. In terms of the free
fields they are given by

$$
s_1^{(2)}=-\frac{1}{2i\pi}N_j\int_{C}\,dt(-t)^{2j-s-1}(\l{-1})^{p-\nu}
(1+t\o{0})^{-2j-2}e^{2i\alpha_0(j+1)x_0}\vac\,,
$$
$$
s_2^{(2)}=-\frac{1}{2i\pi}N_j^{\prime}\int_{C}\,dt(-t)^{-s-2}(1+t\o{0})^{-2j-2}
e^{2i\alpha_0(j+1)x_0}\vac\,,
$$
$$
s_3^{(2)}=-\frac{1}{2i\pi}N_j^{\prime}\int_{C}\,dt(-t)^{-s-2}(\l{-1})^{p+\nu}
(1+t\o{0})^{-2j-2}e^{2i\alpha_0(j+1)x_0}\vac\,,
$$
$$
s_4^{(2)}=-\frac{1}{2i\pi}N_j\int_{C}\,dt(-t)^{2j-s-1}
\left(\j-{0}(\o{n},\l{n},\a{n})\right)^{2p-\nu}
(1+t\o{0})^{-2j-2}e^{2i\alpha_0(j+1)x_0}\vac\,.
$$
where $N_j^{\prime}=\Gamma(2j+2)/(s+2)\Gamma^{1/2}(1+s-2j)$.

Their norms in the case s=-1 are given by

$$
\langle s_1|s_1\rangle ^{(2)}= (p-\nu+1)\,,\qquad \langle s_2|s_2\rangle
^{(2)}=
\langle s_3|s_3\rangle ^{(2)}=\langle s_4|s_4\rangle ^{(2)}=0\,.
\eqno{(5.12)}
$$

In the above, we consider the $j=0$ case for the vector $s_4^{(2)}$ because its
explicit expression in terms of the free fields has a rather complicated
form. Recall that the same is in the case of the vector $s_2^{(1)}$ for
the Dotsenko representation (see sec.4.1). In this sense the $2$-representation
has simpler expressions for the primary fields and singular vectors than
the $1$-representation. However the generalization to a general $j$ is
straightforward.

 From (5.12) it follows that the structure of $V_j^{(2)}$ is described by the
diagram shown in fig.10. Finally, the structure of the $2$-module is given by
the diagram presented in
fig.11.

\vspace{1cm}
\unitlength=1mm
\special{em:linewidth 0.4pt}
\linethickness{0.4pt}
\begin{picture}(40.00,40.67)
\put(20.00,20.00){\circle*{1.33}}
\put(35.00,40.00){\circle*{1.33}}
\put(33.00,37.67){\vector(-3,-4){11.67}}
\put(42.00,40.00){\makebox(0,0)[cc]{$|\nu\rangle^{(2)}$}}
\put(15.00,20.00){\makebox(0,0)[cc]{$s_1^{(2)}$}}
\end{picture}

\vspace{-1.5cm}
\hspace{-0.8cm}
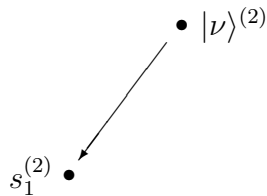
Fig.10: The submodule structure of $V_j^{(2)}$.
\vspace{0.5cm}

\vspace{-4cm}
\unitlength=1mm
\special{em:linewidth 0.4pt}
\linethickness{0.4pt}
\begin{picture}(105.00,0.67)
\put(85.00,-80.00){\circle*{1.33}}
\put(85.00,-60.00){\circle*{1.33}}
\put(85.33,-40.00){\circle*{1.33}}
\put(100.00,-70.00){\circle*{1.33}}
\put(100.00,-51.50){\circle*{1.33}}
\put(100.00,-30.00){\circle*{1.33}}
\put(100.00,-10.00){\circle*{1.33}}
\put(100.00,0.00){\circle*{1.33}}
\put(85.00,-20.00){\circle*{1.33}}
\put(100.00,-28.00){\vector(0,1){15.67}}
\put(100.00,-33.67){\vector(0,-1){15.67}}
\put(100.00,-68.33){\vector(0,1){15.00}}
\put(85.00,-63.33){\vector(0,-1){15.00}}
\put(85.00,-58.33){\vector(0,1){15.33}}
\put(98.33,-3.67){\vector(-3,-4){11.67}}
\put(97.67,-14.33){\vector(-1,-2){11.67}}
\put(97.67,-35.67){\vector(-1,-2){11.67}}
\put(97.67,-54.67){\vector(-1,-2){11.67}}
\put(97.33,-49.00){\vector(-4,3){9.67}}
\put(97.33,-69.33){\vector(-4,3){10.67}}
\put(107.00,0.00){\makebox(0,0)[cc]{$|\nu\rangle^{(2)}$}}
\put(110.00,-10.00){\makebox(0,0)[cc]{$|-\nu\rangle^{(0)}$}}
\put(105.00,-30.00){\makebox(0,0)[cc]{$c_1^{(0)}$}}
\put(106.00,-50.00){\makebox(0,0)[cc]{$m_1^{(0)}$}}
\put(105.00,-70.00){\makebox(0,0)[cc]{$c_2^{(0)}$}}
\put(80.00,-80.00){\makebox(0,0)[cc]{$s_2^{(0)}$}}
\put(80.00,-60.00){\makebox(0,0)[cc]{$m_2^{(0)}$}}
\put(80.00,-40.00){\makebox(0,0)[cc]{$s_1^{(0)}$}}
\put(80.00,-20.00){\makebox(0,0)[cc]{$s_1^{(2)}$}}
\put(27.00,-57.00){\makebox(0,0)[cc]{Fig.11: The submodule structure of}}
\put(26.00,-63.00){\makebox(0,0)[cc]{the $2$-module $\F{j}{2}$.}}
\end{picture}
\vspace{5cm}
\newpage
Note that the $2$-module contains the Verma module part at the top and the
Wakimoto part at the bottom.

\vspace{0.5cm}

\begin{center}
{\rm 5.2 BRST-LIKE COMPLEX OF 2-MODULES}
\end{center}

\vspace{0.5cm}

Let us now construct BRST-like complex of $2$-modules.

We first define the modules with the positive ghost numbers. They have the same
structure as shown in fig.11.

Set

$$
\f{2}{b}=\left\{
\begin{array}{ll}
\F{j+pn}{2}\,,\qquad & b=2n\,,\qquad n\in Z_+ \\
\F{-j-1+pn}{2}\,,& b=2n+1\,,
\end{array} \right.
\eqno{(5.13)}
$$
where $\f{2}{0}\equiv \F{j}{2}$. The number $b$ is called the ghost number, in
analogy to gauge theories.

Now let us turn to the construction of the $2$-modules with the negative ghost
numbers. The starting point is the Wakimoto modules $\f{0}{g}$ with $g<1$.
Define a space $\f{2}{b}$ as $\f{2}{b}=\f{0}{b+1}/S\f{0}{b+1}$, where
$S\f{0}{b+1}$ is a submodule generated by the vector $\hv{\nu+pb}{0}$, if
$b=2n$, or by $\hv{\nu+p(b-1)}{0}$, if $b=2n-1$.
By construction of $\f{2}{b}$ one has the arrows of the diagram shown in
fig.12.
\vspace{1cm}

\unitlength=1mm
\special{em:linewidth 0.4pt}
\linethickness{0.4pt}
\begin{picture}(86.00,85.67)
\put(65.00,45.00){\circle*{1.33}}
\put(65.33,65.00){\circle*{0.00}}
\put(65.33,65.00){\circle*{1.33}}
\put(80.00,75.00){\circle*{1.33}}
\put(80.00,85.00){\circle*{1.33}}
\put(80.00,55.00){\circle*{1.33}}
\put(80.00,35.00){\circle*{1.33}}
\put(65.00,25.00){\circle*{1.33}}
\put(80.00,52.67){\vector(0,-1){15.00}}
\put(80.00,57.33){\vector(0,1){15.33}}
\put(80.00,83.00){\vector(0,-1){6.00}}
\put(65.33,62.00){\vector(0,-1){14.67}}
\put(65.00,27.33){\vector(0,1){15.33}}
\put(78.00,82.33){\vector(-3,-4){11.33}}
\put(77.67,72.33){\vector(-1,-2){11.00}}
\put(78.00,51.67){\vector(-1,-2){11.33}}
\put(77.67,35.67){\vector(-4,3){10.33}}
\put(77.33,56.00){\vector(-3,2){10.00}}
\put(60.00,65.00){\makebox(0,0)[cc]{$m_2^{(0)}$}}
\put(60.00,45.00){\makebox(0,0)[cc]{$s_2^{(0)}$}}
\put(60.00,25.00){\makebox(0,0)[cc]{$m_4^{(0)}$}}
\put(86.00,35.00){\makebox(0,0)[cc]{$m_3^{(0)}$}}
\put(86.00,55.00){\makebox(0,0)[cc]{$c_2^{(0)}$}}
\put(86.00,75.00){\makebox(0,0)[cc]{$m_1^{(0)}$}}
\put(86.00,85.00){\makebox(0,0)[cc]{$c_1^{(0)}$}}
\end{picture}
\vspace{-1.1cm}
\begin{center}
Fig.12: The submodule structure of the $2$-module $\f{2}{b}\,\,,\,b<0$.
\end{center}

{\bf Proposition 5.2}

\vspace{0.3cm}
\hspace{-0.8cm}
{\it The following infinite sequence}

$$
{}\stackrel{\q{2}{-2}}{\rightarrow}\f{2}{-1}\stackrel{\q{2}{-1}}{\rightarrow}
\f{2}{0}\stackrel{\q{2}{0}}{\rightarrow}\f{2}{1}\stackrel{\q{2}{1}}{\rightarrow}
\f{2}{2}\stackrel{\q{2}{2}}{\rightarrow}\,,
$$
$$
\q{2}{b+1}\q{2}{b}=0\,,
$$
$$
\q{2}{b}=\left\{
\begin{array}{ll}
Q_{p-\nu} \,,\quad & b=2n\,,\qquad n\in Z \\
Q_{\nu}\,,& b=2n+1\,,
\end{array} \right.
$$
{\it is a complex}\footnote{In fact $\q{2}{b}$ is the BRST operator with
respect
to the Wakimoto part of this complex (see below).}.

We denote this complex by ${\bf F}^{(2)}$. Notice that one has a relation
between the ghost numbers of
${\bf F}^{(0)}$ and ${\bf F}^{(2)}$ complexes.

$$
q=b+1
\eqno{(5.14)}
$$

{\bf Proof}. The proof of this proposition follows from the structures
of the ${\bf F}^{(0)}$ complex of the Wakimoto modules, and of
the Feigin-Fuchs complex of the Verma modules as well as from the
above construction.

The detailed structure of ${\bf F}^{(2)}$ can be pictured as in fig.13.

\vspace{1cm}
\unitlength=1mm
\special{em:linewidth 0.4pt}
\linethickness{0.4pt}
\begin{picture}(149.33,80.00)
\put(70.00,15.00){\circle*{1.33}}
\put(70.00,35.00){\circle*{1.33}}
\put(80.00,25.00){\circle*{1.33}}
\put(80.00,45.00){\circle*{1.33}}
\put(80.00,65.00){\circle*{1.33}}
\put(80.00,75.00){\circle*{1.33}}
\put(70.00,55.00){\circle*{1.33}}
\put(108.00,55.00){\circle*{1.33}}
\put(108.00,45.00){\circle*{1.33}}
\put(98.00,35.00){\circle*{1.33}}
\put(98.00,15.00){\circle*{1.33}}
\put(108.00,25.00){\circle*{1.33}}
\put(126.00,15.00){\circle*{1.33}}
\put(136.00,25.00){\circle*{1.33}}
\put(136.00,35.00){\circle*{1.33}}
\put(52.00,65.00){\circle*{1.33}}
\put(52.00,45.00){\circle*{1.33}}
\put(52.00,25.00){\circle*{1.33}}
\put(42.00,15.00){\circle*{1.33}}
\put(42.00,35.00){\circle*{1.33}}
\put(24.00,45.00){\circle*{1.33}}
\put(24.00,25.00){\circle*{1.33}}
\put(14.00,15.00){\circle*{1.33}}
\put(16.33,15.00){\vector(1,0){24.00}}
\put(72.33,15.00){\vector(1,0){23.67}}
\put(149.33,15.00){\vector(-1,0){20.33}}
\put(110.33,25.00){\vector(1,0){23.67}}
\put(54.00,25.00){\vector(1,0){23.67}}
\put(7.67,25.00){\vector(1,0){14.33}}
\put(45.00,35.00){\vector(1,0){23.00}}
\put(134.00,35.00){\vector(-1,0){33.33}}
\put(82.00,45.00){\vector(1,0){23.67}}
\put(26.00,45.00){\vector(1,0){24.00}}
\put(105.67,55.00){\vector(-1,0){33.00}}
\put(54.00,65.00){\vector(1,0){23.67}}
\put(78.67,72.67){\vector(-1,-2){7.67}}
\put(80.00,47.67){\vector(0,1){14.67}}
\put(80.00,42.67){\vector(0,-1){15.67}}
\put(70.00,17.00){\vector(0,1){15.67}}
\put(78.33,61.00){\vector(-1,-3){8.00}}
\put(78.33,41.33){\vector(-1,-4){6.00}}
\put(77.67,27.33){\vector(-1,1){6.00}}
\put(106.33,52.33){\vector(-1,-2){7.67}}
\put(107.00,42.33){\vector(-1,-3){8.33}}
\put(108.00,27.33){\vector(0,1){14.33}}
\put(134.67,32.67){\vector(-1,-2){7.67}}
\put(52.00,63.00){\vector(0,-1){16.00}}
\put(51.33,62.67){\vector(-1,-3){8.33}}
\put(42.00,32.67){\vector(0,-1){15.67}}
\put(52.00,27.00){\vector(0,1){16.00}}
\put(50.33,42.33){\vector(-1,-4){6.33}}
\put(50.33,26.67){\vector(-1,1){6.67}}
\put(24.00,42.67){\vector(0,-1){15.67}}
\put(23.67,43.00){\vector(-1,-3){8.67}}
\put(24.00,50.00){\makebox(0,0)[cc]{$\f{2}{-2}$}}
\put(52.00,70.00){\makebox(0,0)[cc]{$\f{2}{-1}$}}
\put(80.00,80.00){\makebox(0,0)[cc]{$\f{2}{0}$}}
\put(108.00,60.00){\makebox(0,0)[cc]{$\f{2}{1}$}}
\put(136.00,40.00){\makebox(0,0)[cc]{$\f{2}{2}$}}
\put(140.00,10.00){\makebox(0,0)[cc]{(3)}}
\put(118.33,30.67){\makebox(0,0)[cc]{(2)}}
\put(88.67,51.33){\makebox(0,0)[cc]{(1)}}
\end{picture}
\vspace{-0.5cm}
\begin{center}
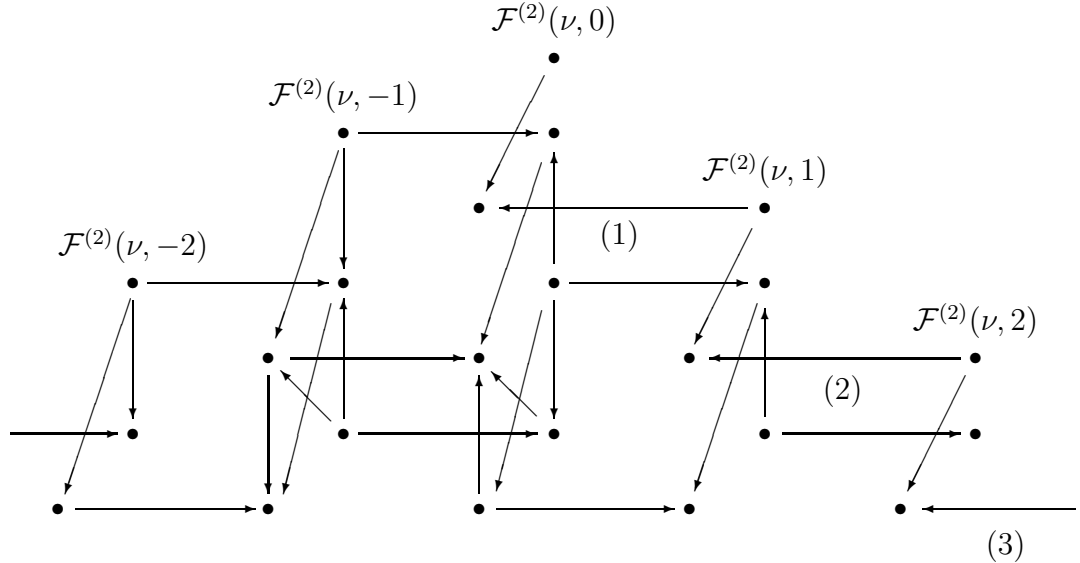
Fig.13: The complex ${\bf F}^{(2)}$. Horizontal arrows indicate where special
vectors are mapped to under the BRST operator. The other arrows indicate the
submodule structure of the $2$-modules.
\end{center}

By construction of ${\bf F}^{(2)}$, the horizontal arrows of its bottom part
are
the ones of the ${\bf F}^{(0)}$ complex (see fig.3). The arrows
(i) of its top part are due to the Feigin-Fuchs complex of the Verma modules
[15]. It is easy to see that the ghost number $b$ is one with respect to the
Wakimoto part of the complex ${\bf F}^{(2)}$. The ghost number for the Verma
part is equal to $-b$. It should be noted that the BRST operator for the Verma
part is not expressed in terms of the free fields.

Finally, one can compute the cohomology of this complex

$$
H^b=\frac{{\rm Ker}\,\q{2}{b}}{{\rm Im}\,\q{2}{b-1}}=\left\{
\begin{array}{ll}
0\,,\qquad & b\neq 0\\
{\cal H}_\nu\,, & b=0\,.
\end{array}\right.
\eqno{(5.15)}
$$
where ${\cal H}_{\nu}$ is the irreducible $\hat{sl}_2$-module (Hilbert space of
the model).

It is expected that there is an infinite set of modules. The Verma and Wakimoto
modules are the boundary modules of this set. Repeating the procedure used in
sec.4 and 5 one can step by step build an arbitrary module from this set. In
fact we have the following diagram

$$
\bullet ^{\infty}\,\cdot\,\cdot\,\cdot\,\bullet ^{2}\longleftarrow\bullet
^{1}\longleftarrow\bullet ^{0}
$$
\vspace{0.3cm}

\hspace{-0.8cm}
where $0$ corresponds to the Wakimoto module, $1$ to the Dotsenko
module,..., $\infty$ to the Verma module.

\vspace{0.5cm}

\begin{center}
{\rm 5.3 CONFORMAL BLOCKS ON THE PLANE}
\end{center}

\vspace{0.5cm}

In this section we use the $2$-representation together with the Wakimoto
representation to give an explicit calculation for conformal blocks on the
plane. As an example we compute the two-point function of the primary fields.

Set $\vc =\vac\,,\cvc =\cvac 1^{(2)}{}^{\dagger}$, where $1^{(2)}$ is the
$2$-representation for the identity operator. The simplest two-point function
is given by\footnote{In this representation as well as in the Wakimoto there
are simple explicit expressions for the fields $\phi ^j_m(z)$ (see
(3.6a),(5.4)).}

$$
\langle \phi ^j_{-m}(z_1)\phi ^j_m(z_2)\rangle \,.
$$

Chose the following representations for the primary fields
$\rpp{-m}{j}{0}{z_1}$
and $\rpp{m}{j}{2}{z_2}$. It is easy to see that one does not need
either additional screening operators or identity operators. In order to
compare the calculation with the one of sec.4.3 we set $m=j$. Then

$$
\langle\phi ^j_{-j}(z_1)\phi ^j_j(z_2)\rangle^{({\vec\alpha},{\vec\beta},
{\vec\gamma},{\vec\lambda})}=\cvac 1^{(2)}{}^{\dagger} \rpp{-j}{j}{0}{z_1}
\rpp{j}{j}{2}{z_2} \vac\,.
\eqno{(5.16)}
$$
Here ${\vec\alpha}=(0,2)\,,{\vec\beta}=(0,2)\,,{\vec\gamma}={\vec\lambda}=0$.

Using (4.10) as well as the formulas (3.6a),(5.4) for the primary
fields and taking the limit $s\rightarrow -1$, one arrives at the result

$$
\langle\phi ^j_{-j}(z_1)\phi ^j_j(z_2)\rangle^{({\vec\alpha},{\vec\beta},
{\vec\gamma},{\vec\lambda})}=C_2/z_{12}^{2\triangle}\,,\qquad
C_2=i^{4j}(2j!)\,.
\eqno{(5.17)}
$$
$C_2$ can be absorbed into a proper normalization of the primary fields.

Let us conclude this section with some remarks.

(i) In contrast to the previous case (4.21) the $z_{12}-dependence$ in (5.17)
is
only due to the scalar field $\varphi$. The fields $(\omega,\omega^{\dagger})$
don't give rise to the dependence on $z_{12}$. It is the same as in the
Dotsenko-Fateev representation of a two-point function for the minimal model.

(ii) An interesting result is that the 2-representation does not have the
screening operator. Explicitly

$$
S^{(2)}=\oint_{C_z}\,dz:J^+(z)\rpp{-1}{-1}{2}{z}:\,\,\,\,\sim\oint_{C}\,dt(-t)^{-2}=0\,.
\eqno{(5.18)}
$$

(iii) One has to be careful using the $2$-representation to compute the
conformal blocks, since the vector $\hv{\nu}{2}$ is the highest weight vector
for the factor space (see sect.5.1). The problem is to take this fact into
account in explicit calculations of the conformal blocks.

(iiii) Following [4,8], one can try to introduce a chiral primary field  as

$$
{}_m\Phi _{\nu\rho}^{\mu}{}^{(2)}\,:\, {\cal H}_\nu\rightarrow  {\cal H}_\mu\,,
\eqno{(5.19)}
$$
where $\nu\,,\mu\,,\rho$ are given by (4.22).

However there are two problems. The first problem is to write the field
${}_m\Phi _{\nu\rho}^{\mu}{}^{(2)}$ in terms of the free fields. The second
is to prove the BRST invariance of this field\footnote{A possible solution of
the first problem is ${}_m\Phi _{\nu\rho}^{\mu}{}^{(2)}(z)=\rpp{m}{j}{2}{z}
\prod_{i=1}^{r}\oint_{C_i} \,dz_i\dom{z_i}e^{2i\alpha_0\vp{z_i}}$
(see eq.(4.23)). The second problem is more complicated because an explicit
expression for a BRST operator in the Verma part of ${\bf F}^{(2)}$ is not
available in terms of the free fields.}.

\section{Conclusions}

First let us say a few words about the results.

The free field representation provides in principle a powerful way to obtain
the
correlators and compute the Operator Algebra of the primary fields. In this
paper we have reworked and completed the free field representation for
so-called
degenerate conformal field theories, relying on the singular vectors in the
Verma modules. Based on this construction, the full set of representations for
the primary operators as well as a proper definition for conformal blocks
have been proposed. As an example we have considered
\newpage
the new modules for the
$SL(2)$ WZW model. They are, in fact, the simplest nontrivial modules in the
full set of bosonized highest weight representations of $\hat {sl}_2$ algebra.
The Wakimoto and Verma modules appear as boundary modules of this set. We have
constructed the BRST-like complexes of these modules. We have also computed the
two-point function of the primary fields by using the new nontrivial
representations.

Let us now conclude by mentioning some open problems.

(i) We have seen that the bosonized theory has the additional quantum number.
This means that there is a hidden symmetry in the theory. The problem is to
understand what underlies this symmetry.

Going one step further, one can consider, as an example, the $SL(2)$ WZW model.
One interesting observation is that in this case the screened vertex operators
(chiral primary field) define the representation of the quantum group $SU(2)_q$
[16]. In fact they are a subset of the full set of the representations for
the screened vertex operators. In the bosonized theory there is an
infinite representation of a quantum group. The problem is to understand what
is
a quantum group as well as an algebra of the screening operators\footnote{As we
have seen in above there is an infinite set of the screening operators
$S^{(\gamma)}$ in this theory.}. Note that the quantum group appears due to the
bosonization.

(ii) The second problem is impressive too. It concerns a new type of the
bosonization. The point is that in the well-known Friedan-Martinec-Shenker
bosonization when one writes an initial set of the free fields in terms of a
new
set then the structure of the Fock space changes automatically [17]. In
our bosonization we have the same set of the free fields but we change the
structure of the Fock space by using a different representation\footnote{These
representations correspond to the different values of the new quantum number.}.
Hence, we will call this bosonization as the bosonization {\rm II}.

It would be very interesting to obtain identities between the special functions
by using our bosonization. It seems that the computation of the
characters (partition functions of the irreducible representation ${\cal
H_{\nu}}$ for $\hat{sl}_2$) is the simplest example. More complicated examples
are the correlation functions on the higher genus Riemann surfaces.

(iii) As we have seen in sec.4,5 the $SL(2)$ WZW model yields an example of
a $\beta\gamma$ system with a background charge. It seems that such systems
appear naturally in nontrivial representations and it would be usefully to
develop a theory for these systems. Some steps in this direction have been
taken
in works [10,18].

(iiii) One can use the different representations for the primary fields in
order
to obtain solutions of the Knizhnik-Zamolodchikov equation via the Ward
identities [14]\footnote{In fact, one does not need the bosonized theory in
this case. We can use the expressions for the primary fields in terms of
currents (Malikov-Feigin-Fuchs solutions).}. The generalization is to use more
than one field in nontrivial representation. Then we may find a set of
solutions
via the Ward identities. Now the problem is to separate the basis
(linear-independent solutions) from this set.

(iiiii) One has seen in the example of the $SL(2)$ WZW model that the BRST-like
operators are built via the $S^{(0)}$ screening operator. It is expected that
these operators can also be constructed in terms of an arbitrary screening
operator $S^{(\gamma)}$. From this point of view one has an infinite set of the
BRST-like operators as well as complexes corresponding to them.

(iiiiii) In conclusion, we would like to stress that one can apply the
technique
developed in sec.2 to the minimal models, $W$ algebras, $d=2$ gravity etc. Now
the problem is to find explicit expressions for the singular vectors in these
theories. However one can try to use a such effective tool as
the Drinfeld-Sokolov reduction, which relates $W$ algebras to the Kac-Moody
algebras, in order to overcome this obstacle.

\begin{center}
Acknowledgments
\end{center}
The details of the representations described above have been reported at the
{\rm XXVII} International Symposium Ahrenshoop on the Theory of Elementary
Particles, Wendisch-Rietz, September 1993 and at the 120th WE-Heraeus-Seminar
on
Mathematical Problems of Quantum Theory and Statistical Physics, Bad Honnef,
October 1993. The efforts of the organizers of these activities are gratefully
acknowledged. O.A. is also grateful to Vl.Dotsenko, E.Frenkel for comments
and G.Lopes Cardoso for reading the manuscript. O.A. is supported by DFG.

\begin{center}
Appendix A
\end{center}

In this appendix we will compute an integral used in sec.4 in order to
construct the Dotsenko representation.

Let us consider an integral

$$
I(a)=\int_{C_t}\int_{C_v}\,dtdv (tv)^ae^{tv}\,,
\eqno{(A.1)}
$$
where the integration contours $C_t$ and $C_v$ are shown in fig.4, and where
$a$
is a parameter.

Using the formal formula

$$
\oint_{C_t}\,dt(-t)^{-a}e^{-bt}=-2i\pi b^{a-1}/\Gamma (a)
$$
one easily arrives at the result

$$
I(a)=(2i\pi)^2/\Gamma(-a)\,.
\eqno{(A.2)}
$$

The right hand side of (A.2) can be defined for other values of $a$ by the
analytic continuation.
\newpage
\begin{center}
Appendix B
\end{center}

The purpose of this appendix is to compute an integral relevant for the
$2$-representa-tion.

Let

$$
I(a,b)=\left(\frac{\Gamma(a)}{2i\pi}\right)^2\int_{C_t}\int_{C_v}\,dtdv(tv)^{-a}
\cvac (1-v\l{0})^b(1+t\o{0})^b\vac \,,
\eqno{(B.1)}
$$
where the integration contours $C_t\,,C_v$ are represented in fig.4. $a\,,b$
are parameters and $\o{0}\,,\l{0}$ are zero modes of $\om{z}$ and $\dom{z}$.

Taking into account the commutation relation (3.2), one finds

$$
I(a,b)=\left(\frac{\Gamma(a)}{2i\pi}\right)^2\int_{C_t}\int_{C_v}\,dtdv(tv)^{-a}\,
{}_2F_0(-b,-b;tv)\,,
\eqno{(B.2)}
$$
where ${}_2F_0(-b,-b;tv)$ is the degenerate hypergeometric function [19].

Using the integral representation for ${}_2F_0(-b,-b;tv)$, we arrive at
the following formula

$$
I(a,b)=\left(\frac{\Gamma(a)}{2i\pi}\right)^2\lim_{k\rightarrow\infty}\,\frac{1}
{B(-b,k+b)}\int_{C_t}\int_{C_v}\int_{0}^{1}\,dtdvdx(tv)^{-a}\frac{x^{-b-1}
(1-x)^{k+b-1}}{(1-x(1+ktv))^{-b}}\,.
\eqno{(B.3)}
$$

Finally, it follows from (B.7), the definition of the $B$ function and the
relation $\lim_{|z|\rightarrow\infty}z^x\Gamma(z)/\Gamma(z+x)=1$, that the
integral $I(a,b)$ is given by

$$
I(a,b)=\Gamma(a)\Gamma^2(a-b-1)/\Gamma^2(-b)\,.
\eqno{(B.4)}
$$

Now let us calculate one more integral.

Set

$$
I(a,b,k,x)=\int_{C_t}\int_{C_v}\,dtdv (tv)^a(1-x(1+ktv))^b\,.
\eqno{(B.5)}
$$
Using the hypergeometric function and its integral representation, one obtains

$$
I(a,b,k,x)=\frac{(1-x)^b}{B(-b,1+b)}\int_{C_t}\int_{C_v}\int_{0}^{1}\,dtdv
dy\frac{(tv)^ay^{-b-1}(1-y)^b}{1+ktvxy/1-x}\,.
\eqno{(B.6)}
$$
Choose the parameter $a$, so that ${\rm Re}\,\,a<0$. This suggests that the
contour $C_t$ becomes the one shown in fig.5. Now the point $(1-x)/kvxy$ is
inside of the contour.  By using the Cauchy theorem as well as the definition
of
the $B$ function, we finally arrive at the result

$$
I(a,b,k,x)=(2i\pi)^2(-xk)^{-a-1}(1-x)^{a+b+1}\frac{\Gamma(-a-b-1)}{\Gamma(-a)
\Gamma(-b)}\,.
\eqno{(B.7)} $$

\end{document}